\documentclass[fleqn,usenatbib]{mnras}

\usepackage[T1]{fontenc}

\DeclareRobustCommand{\VAN}[3]{#2}
\let\VANthebibliography\thebibliography
\def\thebibliography{\DeclareRobustCommand{\VAN}[3]{##3}\VANthebibliography}

\usepackage{graphicx}	
\usepackage{xcolor}
\usepackage{amsmath}	
\usepackage{mathtools}
\usepackage{amssymb}	
\usepackage{cprotect}
\usepackage{soul}
\usepackage{afterpage}
\usepackage{lscape}
\usepackage{breakurl}

\newcommand{\bp}{\mbox{$G_{\rm BP}$}}
\newcommand{\rp}{\mbox{$G_{\rm RP}$}}
\newcommand{\Gabs}{\mbox{$G_{\rm abs}$}}

\newcommand{\Teff}{\mbox{$T_{\mathrm{eff}}$}}
\newcommand{\Msun}{\mbox{$\mathrm{M_{\odot}}$}}
\newcommand{\Pwd}{\mbox{$P_{\mathrm{WD}}$}}

\usepackage{newtxtext,newtxmath}

\title{A catalogue of white dwarfs in Gaia EDR3}
\author[Gentile Fusillo et al.]{N.~P.~Gentile Fusillo,$^{1}$\thanks{E-mail: ngentile@eso.org}, P.-E. ~Tremblay$^{2}$, E. ~Cukanovaite$^{2}$, A.~Vorontseva$^3$, R.~Lallement${^4}$,
\newauthor M.~Hollands$^{2}$, B.~T.~G\"ansicke$^{2}$, K.~B.~Burdge${^5}$, J.~McCleery$^{2}$ and S.~Jordan${^6}$\\
$^{1}$European Southern Observatory, Karl Schwarzschild Stra{\ss}e 2, Garching, 85748, Germany\\
$^{2}$Department of Physics, University of Warwick, Coventry, CV4 7AL, UK\\
$^{3}$Sixfold GmbH, Magirus-Deutz-Stra{\ss}e 16, 89077 Ulm, Germany\\
$^{4}$GEPI (Galaxies-Etoiles-Physique-Instrumentation), Observatoire de Paris,
5 Place Jules Janssen, 92195 Meudon Cedex, France\\
${^5}$ Division of Physics, Mathematics and Astronomy, California Institute of Technology, Pasadena, CA 91125, USA\\
${^6}$ Astronomisches Rechen-Institut, Zentrum f\"ur Astronomie, Universit\"at Heidelberg, M\"onchhofstra{\ss}e 12-14, 69120 Heidelberg, Germany\\}

\date{Accepted XXX. Received YYY; in original form ZZZ}

\pubyear{2021}

\begin{document}
\label{firstpage}
\pagerange{\pageref{firstpage}--\pageref{lastpage}}
\maketitle
\begin{abstract}
We present a catalogue of white dwarf candidates selected from \textit{Gaia} early data release three (EDR3).
We applied several selection criteria in absolute magnitude, colour, and \textit{Gaia} quality flags to  remove objects with unreliable measurements while preserving most stars compatible with the white dwarf locus in the Hertzsprung-Russell diagram. We then used a sample of over 30\,000  spectroscopically confirmed white dwarfs and contaminants from the Sloan Digital Sky Survey (SDSS) to map the distribution of these objects in the \textit{Gaia} absolute magnitude-colour space. 
Finally, we adopt the same method presented in our previous work on \textit{Gaia} DR2 to calculate a probability of being a white dwarf (\Pwd) for $\simeq 1.3$ million sources which passed our quality selection.
The \Pwd\ values can be used to select a sample of $\simeq 359\,000$ high-confidence white dwarf candidates. We calculated stellar parameters (effective temperature, surface gravity, and mass) for all these stars by fitting \textit{Gaia} astrometry and photometry with synthetic pure-H, pure-He and mixed H-He atmospheric models.
We estimate an upper limit of 93~per cent for the overall completeness of our catalogue for white dwarfs with $G \leq 20$\,mag and effective temperature (\Teff)\ $> 7000$\,K, at high Galactic latitudes ($|b|>20^{\circ}$). Alongside the main catalogue we include a reduced-proper-motion extension containing $\simeq10\,200$ white dwarf candidates with unreliable parallax measurements which could, however be identified on the basis of their proper motion. We also performed a cross-match of our catalogues with SDSS DR16 spectroscopy and provide spectral  classification based on visual inspection for all resulting matches. \end{abstract}




\section{Introduction}
White dwarfs are by far the most common stellar remnants in the Galaxy and over 95~per cent of all stars will end their lives as one of these small fading embers \citep{fontaineetal01-1}. Several unique properties of white dwarfs make them powerful tools with applications in various areas of astronomy: from flux calibration (e.g. \citealt{bohlinetal14-1}) to cosmochronology (e.g. \citealt{fontaineetal01-1}) and exo-planetary science (e.g. \citealt{hollandsetal18-1}). 
However, the intrinsic low luminosity of  white dwarfs has always posed a significant observational challenge and large, well-defined samples of these stars have historically been difficult to assemble.

In 2018 the second data release of \textit{Gaia} (DR2) led to a true revolution in the field of white dwarf science, with accurate parallax measurements unlocking the possibility to search for these stellar remnants on an unprecedented scale. \citet{jimenez-esteban18} identified $\simeq73\,000$ white dwarfs and explored in more details the population within the 100\,pc solar neighbourhood, and 
\citet{gentilefusilloetal19-1} sampled the entirety of  \textit{Gaia}\,DR2 identifying a total of 
$\simeq260,000$ white dwarfs, an eight-fold increment compared to the number of objects known before \textit{Gaia} \citep{gentilefusilloetal19-1}. 
This new, well-defined and homogeneous sample of white dwarfs gave astronomers an unprecedented opportunity to look at the global properties of these stars, resulting already in a number of important new discoveries. 

\citet{tremblay-nature} identified a `transversal' sequence in the Hertzsprung-Russell (H-R)  diagram of \textit{Gaia} white dwarfs not aligned with theoretical cooling tracks and not explained by a unique atmospheric composition. \citet{tremblay-nature} recognized this feature as the first direct observational evidence of a delay in white dwarf cooling due to core crystallization and associated physics such as phase separation and sedimentation, a feature of the H-R diagram which had been predicted over 50 years before \citep{vanhorne68-1}. \cite{cheng2019} later demonstrated that about 6~per cent of high-mass white dwarfs ($M>1.05\,\Msun$) on this transverse sequence, likely the products of double-degenerate mergers, must experience an extra 8\,Gyr cooling delay not explained by core-crystallization alone. More recently, \citet{blouin21} reconciled these results showing that a distillation process during $^{22}$Ne phase separation 
in crystallising white dwarfs 
could explain both the cooling delay of standard white dwarfs  and the extra delay experienced by high-mass double white dwarf mergers \citep[see also][]{bauer20,cami21}. A number of additional studies have focused on the spectral properties of ultra-massive white dwarfs, consolidating the idea that many of these systems are the result of double white dwarf mergers \citep{hollandsetal20-1,kawkaetal20-1,kilicetal21-1}.

In addition to enabling a close look at the H-R diagram of white dwarfs, the parallax measurements of \textit{Gaia} allowed to more precisely estimate white dwarf fundamental parameters and also calculate them independently of spectroscopy. Consequently, in the wake of DR2 a number of studies  revisited the stellar parameters of various subsets of white dwarfs, evaluated potential systematic offsets in the data of \textit{Gaia} and of various additional large-area surveys, and provided a new statistical view on the global properties of white dwarfs \citep[see, e.g.,][]{tremblay2019, bergeronetal19-1, coutu19,ouriqueetal19-1, chandraetal20-1}. The white dwarf luminosity function was also re-explored with unprecedented level of detail \citep{torresetal21-1}; and \citet{torresetal19-1} further investigated the memberships of white dwarfs into the thin disc, thick disc and halo Galactic populations. Significant progress was also made for large scale identification and characterisation of white dwarfs in binaries with main sequence stars, either in common proper motion pairs \citep{el-badryetal18-1}, non-interacting unresolved systems \citep{inightetal21-1},  and cataclysmic variables \citep{palaetal20-1,abriletal20-1}.

In addition to providing new insight into the global properties of white dwarfs, the huge number of new objects discovered thanks to \textit{Gaia} opened-up the opportunity to identify some of the most peculiar and rare types of white dwarfs.

For example, \citet{kaiseretal21-1} and  \citet{hollandsetal21-1} discovered five cool ($\Teff<5000$\,K) white dwarfs with trace Li in their atmospheres. This rare polluting element is extremely difficult to detect in hotter white dwarfs and could be the signpost of accretion of the crust of a planetary object \citep{hollandsetal21-1}. 
More discoveries related to planetary systems around white dwarfs enabled by \textit{Gaia} included: 
WD\,J0914+1914, a peculiar white dwarf in the process of evaporating a Neptune-like exo-planet \citep{gaensickeetal19-1}; 
and the 14 newly identified  white dwarfs with gaseous debris from rocky planetesimals \citep{melisetal20-1,dennihyetal20-2,gentilefusilloetal21-1}, which brought the number of such systems known from seven to 21. 

However, despite the enormous progress based on the analysis of the \textit{Gaia} DR2 white dwarf samples, they are not without limitations. We estimated the \citet{gentilefusilloetal19-1} catalogue to be mostly complete only
out to $\simeq70$\,pc, but even within the 20\,pc Solar neighbourhood a handful of historically known white dwarfs did not have reliable \textit{Gaia} observations. Furthermore the coolest and therefore reddest white dwarfs remained difficult to be systematically identified both because of their low luminosity and because  of relatively high contamination from other red sources with spurious \textit{Gaia} measurements. 

The early Data Release 3 of \textit{Gaia} (EDR3) relies on 34 months of observations (compared to 22 months for DR2) and represents an improvement on all fronts over DR2, with parallax measurements being now on average 20 to 30~per cent more accurate and proper motion measurements twice as accurate as in the previous DR. Additionally, EDR3 includes new flags and diagnostic parameters which allow to better assess the data available for each source and make more robust quality cuts \citep{Gaia_summary20-1,lindegrenetal20-1, rielloetal20-1}.

\section{Identifying white dwarfs in \texorpdfstring{\textit{Gaia}}{Gaia} EDR3}\label{main_cut}
\subsection{EDR3 quality filtering}
\begin{figure*}
\includegraphics[width=2.\columnwidth]{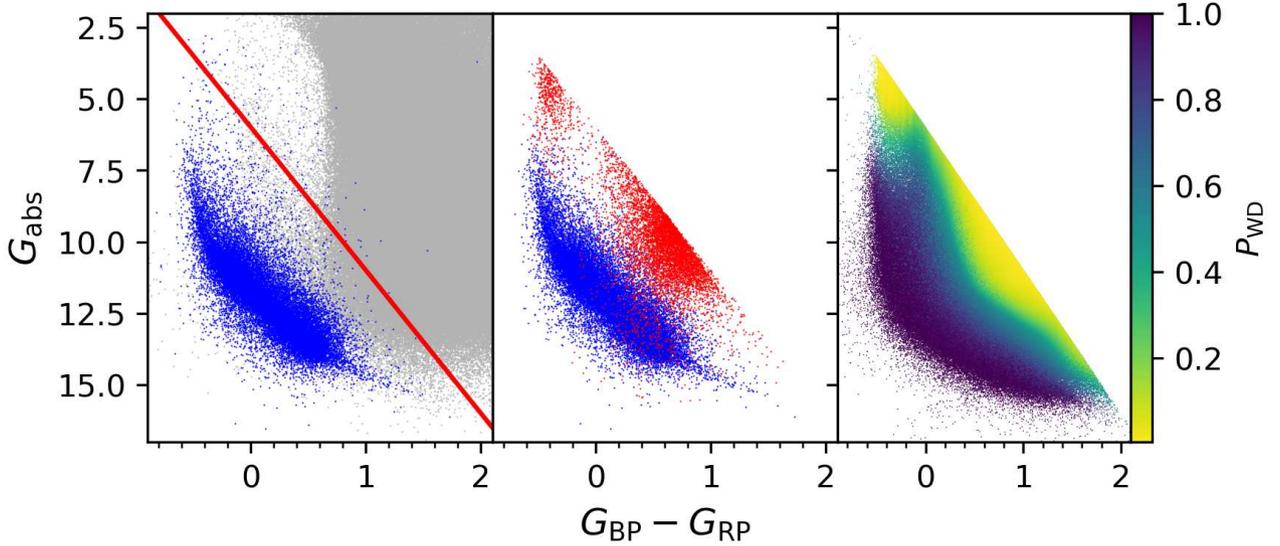}
\caption{\label{maincat_plot} \textit{Left panel:} \textit{Gaia} H-R diagram showing a representative sample of 2 million objects (randomly picked using their $\textsc{random\_index}$) with $\textsc{parallax\_over\_error} >1$ (gray points). The blue points represent the SDSS DR16 spectroscopically confirmed white dwarfs used to broadly define the white dwarf locus. The initial cut adopted for our selection is indicated by the red solid lines. \textit{Center panel}: Distribution of spectroscopically confirmed SDSS white dwarfs (blue) and contaminants (red) included in our final \textit{Gaia} sample. \textit{Right panel:} \textit{Gaia} H-R diagram of all 1\,280\,266 objects in our catalogue. The colour scale reflects the \Pwd\ value of each object.}
\end{figure*}

\begin{table}
\centering
\caption{\label{summary} Summary of the white dwarf candidate selection in \textit{Gaia} EDR3.}
\begin{tabular}{lr}
\hline\\[-1.5ex]
Total number of sources in \textit{Gaia} EDR3 & 1\,811\,709\,771\\
Sources in initial colour-\Gabs~cuts (Eqs.\,1-2) & 14\,422\,222\\
N. objects after quality filtering (Eqs.\,3-21) & 1\,280\,266 \\
High-confidence candidates ($\Pwd>0.75$) & 359\,073\\
\hspace{0.5cm}of which with $G\leq16$ & 2034\\
\hspace{0.5cm}of which with $16<G\leq18$ & 20\,973\\
\hspace{0.5cm}of which with $18<G\leq20$ & 188\,784\\
\hspace{0.5cm}of which with $G>20$ & 147\,282\\
\\
N. objects in RPM-extension (Sect.\,\ref{sect_ext}) & 113\,572\\
    \hspace{0.5cm}of which with $P_{\mathrm{HWD}}>0.85$ & 10\,200\\
\hline
\end{tabular}
\end{table}

The procedure we employed to select white dwarfs in  \textit{Gaia} EDR3 is in many aspects analogous to the one we developed for \cite{gentilefusilloetal19-1}, but we do not directly rely on any result from our previous work on DR2. As advised in \citet{fabriciusetal20-1} the EDR3 data-set should be considered independent of DR2 and, therefore, we carried out our selection entirely anew.
We began by retrieving EDR3 photometry and astrometry for $\simeq128\,000$ objects with available spectroscopy in the Sloan Digital Sky Survey data release 16 (SDSS\,DR16) with $u-g$, $g-r$ colours consistent with those of white dwarfs and with  \textsc{parallax\_over\_error} $>1$. 
We visually inspected these spectra and reliably identified a total of 25\,655 white dwarfs which we used to visualize the full extent of the white dwarf locus in the \textit{Gaia} H-R diagram (Fig.\,\ref{maincat_plot}). We then implemented a broad cut which defines the area in H-R space within which all white dwarfs with reliable EDR3 measurements are expected to be found (Eq. 1), and limits the number of objects to which all subsequent steps in our selection are applied.
\begin{flalign}
&\Gabs  >6 + 5 \times (\bp-\rp) \\
&\textsc{and~} \textsc{parallax\_over\_error} >1
\end{flalign}
It is important to notice that the white dwarf locus defined in this way can only be considered fully inclusive for single white dwarfs, double white dwarf binaries and white dwarfs with low luminosity companions which do no significantly contribute to the \textit{Gaia} colour. For the rest of the paper all mentions of white dwarfs refer only to this type of systems.
Some white dwarfs with unresolved main sequence companions are also included by our initial selection, but the full parameter space spanned by this type of binaries is considerably larger (see figures 3, 6, 8 and 15 in \citealt{inightetal21-1}) and a significant fraction of these systems cannot be identified using only \textit{Gaia} data. Eqs.\,1-2 provide a broad definition of the white dwarf locus, but they include a total of 14\,422\,222 sources, a large fraction of which  have  unreliable photometric and astrometric measurements and need to be filtered out. The quality filtering criteria used in \citet[\textsc{RUWE} $<1.4$ and \textsc{ipd\_frac\_multi\_peak} $\leq2$ and \textsc{ipd\_gof\_harmonic\_amplitude} $<0.1$]{fabriciusetal20-1} 
only remove 20~per cent of the objects in this sample, but, at the same time, exclude $\simeq11$~per cent of SDSS spectroscopically confirmed white dwarfs with $G<20$ and, therefore, on their own are inadequate for our final aim.  In order to maximize the completeness of our final catalogue we defined a series of quality cuts using a combination of several EDR3 parameters. Using our spectroscopic sample as a reference, this selection aims to remove the vast majority of contaminants sources while preserving all stars which genuinely belong in the white dwarf locus. 
 
However, no unique set of quality criteria can be applied uniformly to the entire sky. Crowded areas remain more challenging even in EDR3 and quality cuts which produce relatively clean samples in low crowding regions do not produce equally good results in more densely populated parts of the sky. Therefore, stricter selection criteria need to be applied for stars in these locations. In order to efficiently deal with this problem, we split the entire EDR3 sample in bins of $\simeq50$ arcsec$^2$, counted the objects within each bin and assigned all \textit{Gaia} sources a \textsc{density} parameter defined as the total number of objects in its bin. This value can then be used to define a threshold beyond which stricter selection criteria are required.
Additionally we divided the sky in three main areas within which we carried out our quality filtering separately: High galactic latitudes, Galactic plane, and Magellanic Clouds. 
The Magellanic Clouds area was defined as a $15^{\circ}$ radius around $\alpha=81.28^{\circ}, \delta=-69.78^{\circ}$ (for the Large Magellanic Cloud) plus a $9^{\circ}$ radius around $\alpha=12.80^{\circ}, \delta=-73.15^{\circ}$ (for the Small Magellanic Cloud, \citealt{gaia-collaboration20-1}). 

Sources with \textsc{density} $<400$ even within the Magellanic Clouds or Galactic plane areas were treated analogously to sources in the High Galactic latitude sample.
The final selection criteria adopted are reported in Eqs.\,3-21.
\begin{table}
\normalsize
\begin{tabular}{l|r}
\multicolumn{1}{l}{\textbf{High Galactic latitudes}}\\
\\
(|b| > 25 \textsc{~or~} \textsc{density}$^{\dagger}$ $\leq$ 400) & (3)\\
\rule{0pt}{1.5em}\textsc{and~}\textsc{astrometric\_sigma5d\_max} < 1.5 & (4)\\ 
\multicolumn{2}{l}{\textsc{or~} (\textsc{ruwe} $\leq$ 1.1 \textsc{and~} \textsc{ipd\_gof\_harmonic\_amplitude} < 1)}\\
\rule{0pt}{1.5em}  \textsc{and~((}\textsc{phot\_bp\_n\_obs} > 2  
\textsc{and~}\textsc{phot\_rp\_n\_obs} > 2) & (5)\\
\multicolumn{1}{c}{\textsc{or~}\textsc{phot\_g\_mean\_mag} < 19)} \\
\rule{0pt}{1.5em}
\textsc{and~} (\textsc{parallax\_over\_error} $\geq$ 4  
\textsc{or~} (\textsc{pm}/\textsc{pm\_err}$^{\dagger}$) > 10) & (6)\\
\rule{0pt}{1.5em}
\textsc{and~} |\textsc{phot\_bp\_rp\_excess\_factor\_corrected}$^{\dagger}$| < 0.6 &(7)\\
\rule{0pt}{1.5em}
\textsc{and~}((\textsc{astrometric\_excess\_noise\_sig} < 2 & (8)\\ 
\multicolumn{1}{c}{\textsc{or~}(\textsc{astrometric\_excess\_noise\_sig} $\geq$ 2}\\
\multicolumn{1}{c}{\textsc{and~} \textsc{astrometric\_excess\_noise} < 1.5))}\\ 
\multicolumn{1}{c}{\textsc{or~} \textsc{astrometric\_params\_solved} < 32)}\\
\\

\multicolumn{1}{l}{\textbf{Galactic plane}}\\
\\
(|b| $\leq 25$ \textsc{~and~} \textsc{density}$^{\dagger}$ $>$ 400) & (9)\\
\rule{0pt}{1.5em}
\textsc{and~}\textsc{astrometric\_sigma5d\_max} < 1.5 & (10)\\ 
\multicolumn{2}{l}{\textsc{or~} (\textsc{ruwe} $\leq$ 1.1 \textsc{and~} \textsc{ipd\_gof\_harmonic\_amplitude} < 1)}\\
\rule{0pt}{1.5em}
\textsc{and~((}\textsc{phot\_bp\_n\_obs} > 2  
\textsc{and~}\textsc{phot\_rp\_n\_obs} > 2) & (11)\\
\multicolumn{1}{c}{\textsc{or~}\textsc{phot\_g\_mean\_mag} < 19)} \\
\rule{0pt}{1.5em}
\textsc{and~} (\textsc{parallax\_over\_error} $\geq$ 4  
\textsc{or~} (\textsc{pm}/\textsc{pm\_err}$^{\dagger}$) > 10) & (12)\\
\rule{0pt}{1.5em} \textsc{and~} |\textsc{phot\_bp\_rp\_excess\_factor\_corrected}$^{\dagger}$| < 0.6 &(13)\\
\\
\textsc{and~}((\textsc{astrometric\_excess\_noise\_sig} < 2 & (14)\\ 
\multicolumn{1}{c}{\textsc{or~}(\textsc{astrometric\_excess\_noise\_sig} $\geq$ 2}\\
\multicolumn{1}{c}{\textsc{and~} \textsc{astrometric\_excess\_noise} < 1.5))}\\ 
\multicolumn{1}{c}{\textsc{or~} \textsc{astrometric\_params\_solved} < 32)}\\
\rule{0pt}{1.5em}
\textsc{and~} ((|\textsc{phot\_bp\_rp\_excess\_factor\_corrected}$^{\dagger}$| < &(15)\\
\multicolumn{1}{c}{$5\times$ \textsc{sigma\_excess\_factor}$^{\dagger}$)}\\
\multicolumn{1}{c}{\textsc{or~}(\textsc{parallax\_over\_error} $\geq$ 4}\\
\multicolumn{1}{c}{\textsc{and~} \textsc{astrometric\_sigma5d\_max} $\leq$ 1))}\\
\\
    \multicolumn{1}{l}{\textbf{Magellanic clouds}}\\
    \\
    \textsc{density}$^{\dagger}$ > 400 &(16)\\
\rule{0pt}{1.5em}
\textsc{and~}\textsc{astrometric\_sigma5d\_max} < 1.5 &(17)\\ 
\rule{0pt}{1.5em}
\textsc{and~}(\textsc{phot\_bp\_n\_obs} > 2 
 \textsc{and~}\textsc{phot\_rp\_n\_obs} > 2) & (18)\\
 \rule{0pt}{1.5em}
\textsc{and~} (\textsc{parallax\_over\_error} > 6 &(19)\\ 
\multicolumn{2}{c}{\textsc{or~} (\textsc{parallax\_over\_error} > 2 
\textsc{and~} (\textsc{pm}/\textsc{pm\_err}$^{\dagger}$) > 10))} \\
\rule{0pt}{1.5em}
\textsc{and~}(\textsc{astrometric\_excess\_noise\_sig} < 2 & (20)\\ 
 \multicolumn{1}{r}{\textsc{or~}(\textsc{astrometric\_excess\_noise\_sig} $\geq$ 2}\\ 
\multicolumn{1}{r}{\textsc{and~} \textsc{astrometric\_excess\_noise} < 1.5))} \\ 
\rule{0pt}{1.5em}
\textsc{and~} |\textsc{phot\_bp\_rp\_excess\_factor\_corrected}$^{\dagger}$| < &(21)\\
\multicolumn{1}{c}{5$\times$ \textsc{sigma\_excess\_factor}$^{\dagger}$}\\
\\
\multicolumn{2}{l}{$^{\dagger}$ {\small parameter not provided in the official EDR3 archival distribution,}}\\
\multicolumn{2}{l}{{\small details are provided in the text}}\\
\end{tabular}
\end{table}

The parameters with the largest impact on our selection are:

$\textsc{astrometric\_sigma5d\_max}$, the five-dimensional equivalent to the semi-major axis of the \textit{Gaia} position error ellipse and is useful for filtering out cases where one of the five parameters, or some linear combination of several parameters, is particularly bad \citep{gaiaDR2-ArXiV-2}.

\textsc{pm\_over\_err},  the ratio of total proper motion to total proper motion error and, although it is not provided in EDR3 archive, it can be calculated from \textsc{pm}, \textsc{pmra\_error}, \textsc{pmdec\_error}. 

\textsc{phot\_bp\_rp\_excess\_factor\_corrected}, the \textsc{phot\_bp\_rp\_excess\_factor} corrected for $G_{\mathrm{BP}}-G_{\mathrm{RP}}$ colour dependence as described in \citet{rielloetal20-1}. It is not provided in the EDR3 archive and needs to be calculated following the recipe in \citet{rielloetal20-1}, \textsc{python} code for the calculation is available on public repository\footnote{\label{exc_f}\url{https://github.com/agabrown/gaiaedr3-flux-excess-correction}}. This parameter can be used to filter out sources with inconsistent $G$, $G_{\mathrm{BP}}$ and $G_{\mathrm{RP}}$ photometry, which are particularly prominent in crowded regions.  In our selection we make cuts in \textsc{phot\_bp\_rp\_excess\_factor\_corrected} with respect to  \textsc{sigma\_excess\_factor} which is defined as "the $1\sigma$ scatter for a sample of well behaved isolated stellar sources with good quality \textit{Gaia} photometry" (see section 9.4 in \citealt{rielloetal20-1} for full details). \\

The combined result of our quality filtering for the three sky areas is a sample of 1\,280\,266 objects which represents a compromise between removing the majority of sources with non-optimal \textit{Gaia} measurements and preserving all the stars in white dwarf locus. 

\subsection{Probability of being a white dwarf: \texorpdfstring{\Pwd}{Pwd}}
\label{sect_probs}
Even after applying all the quality filtering described the in previous section, when looking at our sample of 1\,280\,266 objects, white dwarfs do not immediately stand out as a sequence clearly distinct from the rest of the sources in the sample. Consequently any attempt to select  white dwarfs with simple cuts in the H-R diagram would result in incomplete and/or contaminated sample. 

To answer this problem, we adopted the same procedure  described in \citet{gentilefusilloetal19-1}, i.e. we rely on our sample of spectroscopically confirmed SDSS white dwarfs and contaminants to calculate \textit{probabilities of being a white dwarf} ($P_{\rm WD}$) for all objects in our \textit{Gaia}\,EDR3 sample. We used a total of 22\,998 spectroscopically confirmed single white dwarfs and 7124 contaminant objects to map their distribution in H-R ($\bp-\rp$, \Gabs) space (Fig.\,\ref{maincat_plot}).  In order to create a smooth map covering the entire space of interest, every object was treated as a 2D Gaussian, the width of which reflects the $\bp-\rp$ and \Gabs\ uncertainties of the object. For objects with good quality spectra (signal-to-noise-ratio S/N $>10$) these Gaussians were normalised so that their volume equals unity, while to reflect the more uncertain classification of objects with low S/N spectra (S/N $<10$) we used a normalization value of 0.5. This results in two continuous smeared-out density maps one for white dwarfs and one for contaminants. A probability map is then created as the ratio of the white dwarf density map to the sum of both density maps. Regions outside our H-R cut (Eq.\,2) where given a fixed probability value of zero.
This map can then be used to calculate the \Pwd\ of any \textit{Gaia} object by integrating the product of its Gaussian distribution in H-R space with the underlying probability map. 

Our \Pwd\ values allow users to select sub-samples of stars flexibly compromising between the desired completeness and acceptable levels of potential contamination. 

\begin{figure}
\includegraphics[width=0.95\columnwidth]{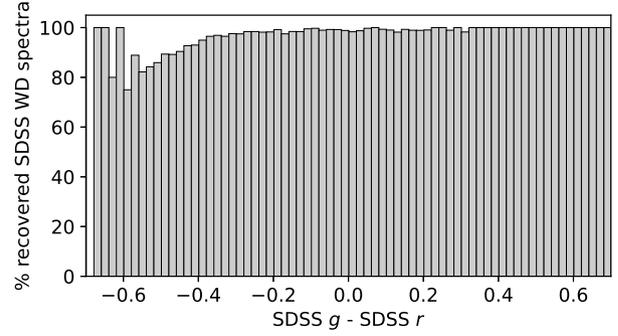}
\caption{\label{colour_hist} Percentage of SDSS white dwarf spectra indentified in our \textit{Gaia} EDR3 catalogue as a function of SDSS $g-r$ colour. The $\simeq20$~per cent drop at $g-r<-0.4$ is likely caused by erroneous inclusion of subdwarfs in the spectroscopic sample.}
\end{figure}

As a generic guideline selecting objects with $\Pwd> 0.75$ recovers $\simeq359\,000$ high-confidence white dwarf candidates, 25\,632 of which have SDSS spectroscopy. $\simeq91$~per cent of these spectroscopic sources are confirmed white dwarfs, $\simeq1$~per cent are contaminant objects, $\simeq3$~per cent are white dwarf-main sequence binaries or cataclysmic variables, and the rest have unreliable classification. When comparing with confirmed SDSS spectroscopic white dwarfs we also find no significant colour bias in
 this selection (Fig.\,\ref{colour_hist}).
 Cleaner, but less complete, white dwarf subsets can be obtained with higher \Pwd\ thresholds and by imposing additional cuts in \textit{Gaia} quality parameters stricter than those already adopted in our selection.

\subsection{The reduced proper motion extension}
\label{sect_ext}
\begin{figure*}
\includegraphics[width=2\columnwidth]{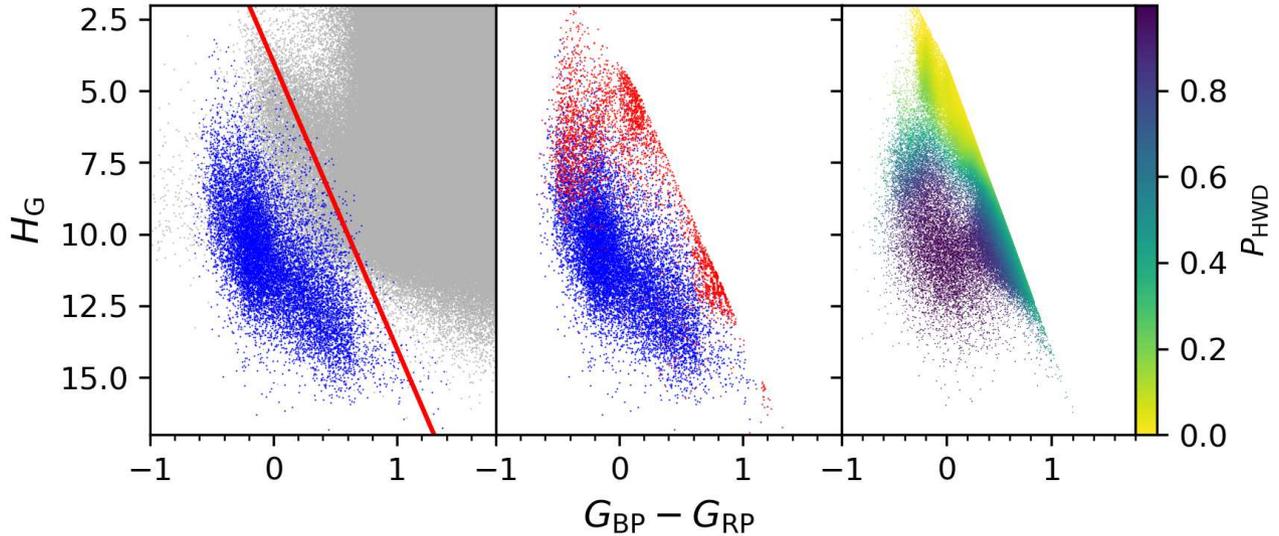}
\caption{\label{Hg_combined} Same as Fig\,\ref{maincat_plot}, but showing the distribution of SDSS spectroscopic sources in reduced proper motion-colour space and the $P_{\mathrm{HWD}}$ distribution for sources in the RPM-extension.}
\end{figure*}

The location in the HR diagram of all \textit{Gaia} sources  with \textsc{parallax\_over\_error} $\leq 1$ was considered too unreliable to be used to identify potential white dwarf candidates. However, a significant fraction of these rejected objects have reliable proper motion measurements. Indeed compared to DR2, proper motion measurements in EDR3 are, on average,  twice as precise while parallax measurements improved only by 20 to 30~per cent. In the absence of reliable parallax estimates, reduced proper motion (RPM) can be used as a proxy for distance and can be employed to distinguish different stellar populations. In particular, before the advent of \textit{Gaia}, colour--RPM diagrams have historically been  used to efficiently select white dwarf candidates \citep[e.g.,][]{jones72-1,harrisetal06-1,gentilefusilloetal15-1,lametal19-1}.
With the aim to fully exploit the potential of \textit{Gaia} as a resource to identify white dwarfs, we decided to create an extension to our main catalogue which contains white dwarf candidates with unreliable parallax measurements, but that could be identified on the basis of their reduced proper motion.
Similarly to what is described in Sect.\,\ref{main_cut} we calculated reduced proper motion defined as
\begin{equation} \tag{22} H_{G}=\textsc{phot\_g\_mean\_mag}+5\log\mu+5\end{equation} for all spectroscopically confirmed white dwarfs and contaminants in our SDSS sample and used these objects to determine the locus occupied by white dwarfs in $H_{\mathrm{G}}$--\textsc{bp\_rp} space (Fig.\,\ref{Hg_combined}). We then retrieved all \textit{Gaia} sources with $\textsc{parallax\_over\_error} \leq1$ and proceeded to define a set of quality cuts aimed at removing sources with unreliable \textit{Gaia} measurements while preserving objects compatible with the white dwarf locus:

\begin{flalign}
&H_{\mathrm{\rm G}}  >10 + 7 \times (\bp-\rp) \tag{23}\\
&\textsc{and~} \textsc{parallax\_over\_error}\leq1 \tag{24}\\
&\textsc{and~} \textsc{density}^{\dagger}<800\tag{25}\\
&\textsc{and~}\textsc{astrometric\_sigma5d\_max}<1.5\tag{26}\\ 
&\textsc{and~}\textsc{phot\_bp\_n\_obs}>3\tag{27}\\ 
&\textsc{and~}\textsc{phot\_rp\_n\_obs}>3\tag{28}\\
&\textsc{and~}(\textsc{pm}/\textsc{pm\_err}^{\dagger})>10)\tag{29}\\
&\textsc{and~} |\textsc{phot\_bp\_rp\_excess\_factor\_corrected}^{\dagger}| < \tag{30} \\[-0.5em]
&\nonumber \hspace{1cm} 3\times \textsc{sigma\_excess\_factor}^{\dagger}
\end{flalign}

This selection results in a sample of 113\,572 objects.
Analogously to what is described in Sect.\,\ref{sect_probs} we created a probability map using SDSS spectroscopic white dwarfs, though in this instance using colour--RPM space instead of colour--\Gabs~space. We then used this map to calculate RPM-based \textit{probabilities of being a white dwarf} ($P_{\mathrm{HWD}}$) for all objects in our RPM extension. Because they are selected among objects with unreliable parallax measurements, all sources in the RPM-extension have relatively poor \textit{Gaia} parameters and are often very faint ($G>20.5$). Furthermore RPM-based probability maps cannot fully distinguish white dwarfs from hot-subdwarfs resulting in some contamination from this type of stars even for relatively high values of $P_{\mathrm{HWD}}$.
We therefore suggest the use of the RPM-extension only for users interested in the faintest white dwarfs approaching the limit of \textit{Gaia} detection and recommend selecting objects with $P_{\mathrm{HWD}} >0.85$. We estimate that a total of $\simeq 10\,200$ genuine white dwarfs are included in the RPM-extension.
Because of the poorer quality of the \textit{Gaia} parameters for the objects in the RPM-extension, in contrast with the main catalogue, we do not provide extinction estimates (see Sect.\,\ref{extinction}), stellar parameters (see Sect.\,\ref{sect_params}) and \textsc{excess\_flux\_error} values (see Sect.\,\ref{var_sect}) for the white dwarf candidates in this sample.

\section{The white dwarf catalogue}
A full version of the main \textit{Gaia}\,EDR3 catalogue of white dwarf candidates and the RPM-extension presented in the previous sections, can be downloaded from: \url{https://warwick.ac.uk/fac/sci/physics/research/astro/research/catalogues/gaiaedr3_wd_main.fits.gz}\\
\url{https://warwick.ac.uk//fac/sci/physics/research/astro/research/catalogues/gaiaedr3_wd_rpm_ext.fits.gz}

\noindent and will also be made available via the Vizier catalogue access tool.

All stars in our catalogue are given a name according to the convention presented in \citet{gentilefusilloetal19-1}, i.e. WD\,JHHMMSS.SS$\pm$DDMMSS.SS defined as the white dwarf coordinates in IRCS, at equinox 2000 and epoch 2000.
Objects which were included in our DR2 catalogue, and already had  WDJ names, have not been re-named (even though updated proper motions may have altered their epoch 2000 projected coordinates) and kept their denomination from \citet{gentilefusilloetal19-1}.
The full catalogue format contains all the columns available in the main \textit{Gaia}\,EDR3 distribution plus a number of additional ones specific to this work (e.g. \Pwd, see Table\,\ref{table_format}). 
We also include photometric and spectroscopic information from the Sloan Digital Sky Survey (see Sect.\,\ref{SDSS_section}) and data from external work on \textit{Gaia}\,EDR3.
For example, \citet{bailer-jones21-1} estimated distances for 1.47 billion objects in \textit{Gaia}\,EDR3 using a probabilistic approach based on a prior constructed from a three-dimensional Galactic model which included interstellar extinction and the non-uniform magnitude limit of \textit{Gaia}. 
Though these distance estimates are not used in the analysis presented in our paper, they represent a valuable added resource for users of our catalogue. We therefore matched all stars in our sample with the \citet{bailer-jones21-1} catalogue using the unique \textit{Gaia}\,EDR3 \textsc{source\_ID} and provide these distance estimates as additional columns.

\begin{figure}
\includegraphics[width=0.95\columnwidth]{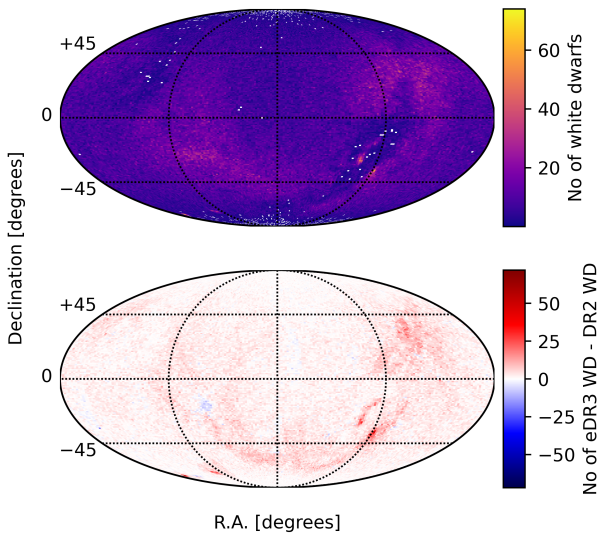}
\caption{\label{all_sky_compare} \textit{Top panel:} Sky density (in 1 deg$^2$ bins) of \textit{Gaia} EDR3 white dwarf candidates with $\Pwd>0.75$ from the catalogue presented in this article. \textit{Bottom panel:} Difference between the number of white dwarf candidates selected in EDR3 (top panel) and the number of DR2 white dwarf candidates with $\Pwd > 0.75$ from \citet{gentilefusilloetal19-1}.}
\end{figure}

\subsection{3D extinction}
\label{extinction}

Extinctions are estimated for each white dwarf in the main catalogue by integrating within new, local 3D maps of differential extinction ($d\leq$ 2.9\,kpc). The differential extinction at a given point P, i.e. the extinction per unit distance along a line of sight crossing P, is proportional to the volume density of the Galactic dust, and is computed by inversion of individual extinctions measured for a large number of target stars distributed in direction and distance. Inversions are based on basic principles described in \cite{Vergely10} and \cite{lallement14} and on recent new, hierarchical techniques presented in \cite{capitanio17} in the frame of the STILISM project and more recently in \cite{lallement19}. Here, we specifically rely on an unpublished map of Galactic interstellar dust that was computed for the EXPLORE project and will be made available at the CDS and online at \url{https://astro.acri-st.fr/gaia_dev/} (Vergely et al., in preparation). The hierarchical inversion is using extinction catalogues from \cite{Sanders18} and \cite{Queiroz20} which are based on major spectroscopic surveys, and, during the inversion, the map presented in \cite{lallement19}, which is based on photometry from \textit{Gaia} and the Two Micron All-Sky Survey (2MASS), is used as a prior distribution and as a final solution in regions of space devoid of spectroscopic survey targets. A more detailed description of this type of data combination, for \cite{Sanders18} data solely,  can be found in \cite{Ivanova21}.  This 4 x 4 x 0.8 kpc$^{3}$ new map (X, Y axes in the Plane and Z axis along Galactic poles) is particularly well suited for nearby stellar sources such as white dwarfs, because spectroscopic surveys are generally targeting brighter, closer stars. We integrated through the new map based on \textit{Gaia} 3D coordinates and resulting parameters are found in columns 115-118 of the catalogue. The extinction $A_{\rm V}$ is given for the standard Johnson $V$ filter and first for the distance corresponding to the exact \textit{Gaia} $1/\varpi$ value. Then we derive the range of extinction corresponding to $1\sigma$ of the parallax value. These estimates neglect uncertainties on the 3D distribution itself and the resulting uncertainty on the integration. Such uncertainties are difficult to quantify individually because they depend on several factors, mainly the target density along the line of sight to the white dwarf, and subsequently the distance, and the uncertainties on the individual extinctions which enter the inversion, but they also depend on the minimum size of clouds reached during the inversion (in our case here 10 pc in most areas). A rough estimate of the average relative error on integrated extinctions is 5~per cent.

In a number of cases the target star is located outside the computational volume of the map and only a lower limit on the extinction is estimated, which corresponds to the extinction reached at the boundary of the map in the direction of the target. Fortunately, in most cases the additional extinction outside the mapped volume is very small, since the dust is concentrated at smaller distance from the Plane than our 400 pc map limit.  In order to quantify the potential additional extinction generated outside this volume, we have used the dust opacity at 353GHz measured by Planck to estimate an upper limit on the total extinction up to infinity. The conversion was done based on the recent determination from \cite{Remy18}, i.e. E(B-V)= 1.5 10$^{4}$ $\tau$353 and the average relationship A$_{V}$=3.1 E(B-V). A flag has been included in the catalog to distinguish among 4 cases, from flag 0 for targets either within the map or outside the map and with potential additional extinction $\delta$ A$_{V}$ $\leq$ 0.2 mag (99.4 percent of objects), to flag 3 for targets outside the map and with potential additional extinction $\delta$ A$_{V}$ $\geq$ 1 mag ($\leq$ 0.03 percent of total).

We do not estimate extinction for the proper-motion extension, as these objects have no reliable distances. While reddening could be estimated for these objects in an iterative fashion by assuming an absolute magnitude using white dwarf models, we consider that this model dependent method is outside the scope of this work. 

The derivation of 3D extinction for all white dwarf candidates in the main catalogue is a considerable astrophysical improvement over the DR2 catalogue of \citet{gentilefusilloetal19-1}, where reddening was instead estimated using a linear distance parameterisation of maximum line of sight extinction from 2D maps \citep{schlegeletal98-1,schlafy}. We find significant changes in extinction for individual objects, largely due to the presence of local interstellar matter fluctuations that were entirely missed from our earlier 2D parameterisation. For white dwarf candidates within 100\,pc the new median extinction is $A_{\rm V}$ = 0.020 mag, with a standard deviation of 0.015 mag, illustrating a fairly homogeneous interstellar medium within this volume. In comparison \citet{gentilefusilloetal19-1} proposed a similar median of $A_{\rm V}$ = 0.027 mag for the same volume, but with a significantly larger and likely unrealistic standard deviation of 0.044 mag.

\section{Stellar parameters}
\label{sect_params}
We derive atmospheric parameters for high-confidence white dwarf candidates using a very similar technique and models as in \citet{gentilefusilloetal19-1}. The main differences compared to our DR2 catalogue, as described in this section, are that we employ EDR3 passbands, use reddening derived from 3D extinction maps, and allow for mixed hydrogen/helium (H/He) compositions in addition to pure-H and pure-He models.

In recent years \textit{Gaia} DR2 photometry has been shown to be reliable for deriving white dwarf fundamental parameters, e.g. when compared to Pan-STARRS, SDSS or J-PLUS \citep{gentilefusilloetal19-1,tremblay2019,bergeronetal19-1,lopez19,mccleery20}, albeit with moderate systematic offsets compared to spectroscopic results \citep{maiz18,tremblay2019,genest19a,narayan19,tremblay20,cukanovaite2021}. \textit{Gaia} EDR3 photometric calibration has already been studied extensively \citep{rielloetal20-1,fabriciusetal20-1} and is equally reliable as that of DR2 (see Sect.\,\ref{40pc}). Possible colour calibration offsets are discussed in Sect.\,\ref{elena}. 

As in the DR2 catalogue we rely exclusively on \textit{Gaia} photometry and astrometry to fit stellar parameters. By construction, reliable \textit{Gaia} data is available for all white dwarf candidates in our catalogue. In contrast, combining \textit{Gaia} data with any other optical photometric survey would require careful quality control, especially for ground-based photometry with significantly different spatial resolution. In addition, the photometric calibration of surveys like SDSS is not fully understood and still relies on ad-hoc corrections partially based on white dwarf spectroscopic parameters \citep{eisensteinetal06-1}. Finally, the lack of constraints on atmospheric composition for most candidates pose a serious challenge to include UV or IR photometric surveys such as \textit{Galex} or \textit{WISE}. While pure-H, pure-He and selected mixed H/He models have been shown to accurately model \textit{Gaia} fluxes in most cases \citep{tremblay2019,bergeronetal19-1,mccleery20}, this conclusion may not be correct for photometry outside of the optical, which could be more sensitive to the H/He ratio or presence of metals. 

There is a degeneracy between $T_{\rm eff}$ and reddening when using optical photometry. We therefore rely on estimated extinction (Sect.\,\ref{extinction}) as an external constraint. We convert extinction to reddening using the standard model of \citet{schlafy}, the same as that used in \citet{gentilefusilloetal19-1}.
\begin{flalign}
A_{G} = 0.835A_{V} \tag{31}\\
A_{G_\mathrm{BP}} = 1.139A_{V} \tag{32}\\
A_{G_\mathrm{RP}} = 0.650A_{V} \tag{33}
\end{flalign}

For sources with six-parameter astrometric solutions, the $G$ band magnitude should be corrected for known systematic offsets \citep{rielloetal20-1}. We applied the correction following the \textsc{python} recipe provided in \citet{Gaia_summary20-1}\footnote{\label{phot_g_corr}\url{https://github.com/agabrown/gaiaedr3-6p-gband-correction}} and used these corrected $G$-band magnitude values to estimate stellar parameters. We provide the corrected magnitude values in our catalogue in the column \textsc{phot\_g\_mean\_mag\_corrected}. The parallax values used for our fits were also adjusted for the known zero point offsets according to the corrections described in \citet{lindegrenetal20-1}\footnote{\label{zp_corr}\url{https://gitlab.com/icc-ub/public/gaiadr3_zeropoint}}. We provide the zero point corrections used in the column \textsc{ZP\_correction}.

We have employed the \textit{Gaia} EDR3 quantum efficiency $S(\lambda)$
for the $G$, $G_{\rm BP}$, and $G_{\rm RP}$ passbands \citep{rielloetal20-1} to calculate
synthetic absolute magnitudes for filter index $i$ using the relation
\begin{flalign}\tag{34}
G^{i}_{\rm abs} = - 2.5 \log \left(\frac{\int S(\lambda)^{i}F (\lambda)\lambda d\lambda}{\int S(\lambda)^{i}\lambda d\lambda} \frac{1}{(3.08568 \times 10^{19}~{\rm cm})^2} \right)
+ C^{i} 
\end{flalign}

{\noindent}where 3.08568 $\times 10^{19}~{\rm cm}$ equates to 10\,pc, and  $C^{i}$ are the zero points derived using \verb|alpha_lyr_stis_002.fits|\footnote{\url{https://archive.stsci.edu/hlsps/reference-atlases/cdbs/calspec/alpha_lyr_stis_002.fits}} as Vega reference spectrum \citep{rielloetal20-1}. The resulting zero points are $C^G$ = $-$21.48503, $C^{\rm BP}$ = $-$20.96683 and $C^{\rm RP}$ = $-$22.22089 mag, respectively. $F(\lambda)$ is the integrated stellar flux in erg s$^{-1}$ \AA$^{-1}$ relatied to the emergent monochromatic Eddington flux $H_{\lambda}$ from model atmospheres as
\begin{flalign}\tag{35}
F(\lambda) = 4\pi R^2 H_{\lambda}({\rm T_{eff}, \log g})
\label{radius}
\end{flalign}
where $R$ is the white dwarf radius and $\log g$ is the surface gravity in cgs units. 

For pure-H and pure-He atmospheres, were rely on the same model atmospheres as in \citet{gentilefusilloetal19-1}. In brief, we use the grid of \citet{tremblayetal11-1} with Lyman $\alpha$ opacity of \citet{kowalski06} for pure-hydrogen composition. The validity range of the grid is $1500 <$ $T_{\rm eff}$ [K] $< 140\,000$ and $6.5 < \log g < 9.5$. For pure-helium composition we use the grid of \citet{bergeron11} in the range of $3500 <$ $T_{\rm eff}$ [K] $< 40\,000$ and $7.0 < \log g < 9.0$. 

A new addition to our catalogue is the use of mixed H/He model atmospheres. In comparison to using pure-He models, a mixed composition of H/He = 10$^{-5}$ has been shown to result in a stable mass distribution for He-rich atmospheres, in much better agreement with the H-rich atmosphere mass distribution as well as predictions from stellar evolution \citep{bergeronetal19-1,mccleery20}. Therefore we employ a third grid of model atmospheres with H/He = 10$^{-5}$ composition based on calculations of \citet{tremblayetal14-1,mccleery20} for the range $2000 <$ $T_{\rm eff}$ [K] $< 40\,000$ and $7.0 < \log g < 9.0$. 

In all cases, the radius in Eq.~\ref{radius} and mass are calculated using evolution sequences from \citet{bedardetal20-1} for thick H-layers (pure-H) or thin H-layers (mixed, pure-He) and $M > 0.46$ $M_{\odot}$. For lower masses, we use the He-core cooling sequences of \citet{serenelli01}. We emphasise that this implies the fitting of only two independent stellar parameters, in our case $T_{\rm eff}$ and $\log g$, with mass and radius being fully determined from the model dependence. 

The dereddened observed \textit{Gaia} flux $f^i$ for passband $i$ in units
of erg cm$^{-2}$ s$^{-1}$ is derived from
\begin{flalign}\tag{36}
G^i = -2.5\log(f^{i}) + C^i
\end{flalign}
{\noindent}which is linked to the passband and stellar disc integrated flux $F^{i}$
in erg s$^{-1}$ as
\begin{flalign}\tag{37}
f^i = \varpi^2 F^i
\end{flalign}
We employ the same fitting technique as in \citet{gentilefusilloetal19-1} based on the non-linear least-squares method
of Levenberg-Marquardt. The uncertainties on stellar parameters are obtained directly from the covariance matrix. Both the uncertainties and reduced $\chi^2$ values are given in our EDR3 catalogue, the latter being useful to flag outliers with colours that deviate from the model grids, such as binaries. 

\subsection{Catalogue stellar parameter values}

The full catalogue includes white dwarf candidates that fall outside of the existing model grids, as well as relatively distant objects for which the parallax uncertainty leads to $T_{\rm eff}$ or $\log g$ error bars that are larger than the full extent of the model grid. Therefore, we adopt the following conditions to have parameters in the catalogue
\begin{flalign}
P_{\rm WD} > 0.70 \tag{38}\\
{\rm AND}~ \sigma_{\rm Teff}/T_{\rm eff} < 0.75 \tag{39}\\
{\rm AND}~ \sigma_{\rm log g} < 2.0 \tag{40}\\
{\rm AND}~ \sigma_{\rm MWD}/M < 1.0 \tag{41}\\
{\rm AND}~ 0.1 < M_{\rm WD}/M < 1.4 \tag{42}\\
{\rm AND}~ 3500 < T_{\rm eff} [K] < 140\,000 \tag{43}
\end{flalign}
with the following additional restrictions for pure-He atmospheres
\begin{flalign}
3500 < T_{\rm eff} [K] < 40\,000 \tag{44}
\end{flalign}
and mixed atmospheres
\begin{flalign}
6600 < T_{\rm eff} [K] < 40\,000  \tag{45}
\end{flalign}
{\noindent}Pure-H and mixed parameters are cut-off at low $T_{\rm eff}$ values compared to the original grid ranges because CIA opacities are likely incorrect in our current grids of models \citep{gentilefusilloetal20-1}. Furthermore, for ultra-cool white dwarfs past the so-called blue-hook in the H-R diagram \citep{hansen98}, there is a degeneracy between cool and hot solutions from Gaia data alone. This can be seen from Fig.~\ref{gcns_diff} where pure-H cooling tracks of mass $M \approx$ 0.60 $M_\odot$ and 2000--3000\,K cross the cooling tracks of more massive white dwarfs at 3000--6000\,K. In our catalogue we always pick the warmer,  massive solution, leading to incorrect parameters for ultra-cool white dwarfs, although only a handful of these objects have so far been identified in the local volume sample \citep{kilic20}. As noted in  \citet{kilic20}, these objects may not be ultra-cool but rather characterised by a peculiar atmospheric composition and faintness in the infrared.

Mass values below 0.2\,\Msun\ and above 1.30\,\Msun\ should be taken with high caution as those parameters were extrapolated outside of the validity range of the available mass-radius relations.  Parameters above 40\,000\,K are extremely sensitive to \textit{Gaia} colours, zero points and reddening corrections and should also be taken with caution without spectroscopic confirmation. Finally, our atmospheric parameters are severely limited for many unresolved binary systems, including double degenerate white dwarfs, although low mass photometric values (over-luminous objects) can still be used to identify promising double degenerate candidates \citep[see, e.g.,][]{bergeron01-1}.

Apart from the limiting cases mentioned above, the catalogued stellar parameters based on state-of-the-art 3D extinction maps are expected to be some of the best available photometric solutions for well behaved DA, DAH, DAZ (pure-H models), DB and DC stars (pure-He or mixed models). We refer to table 2 of \citet{mccleery20} for our proposed choices of solution between pure-H, pure-He or mixed H/He as a function of spectral type and temperature. These parameters can be adopted as precise photometric solutions if the spectral type is already know from other sources, e.g. the SDSS-Gaia catalogue in Sect.\,\ref{SDSS_section} or the Montreal White Dwarf Database \citep{dufouretal17-1}. See Sect.\,\ref{elena} for a comparison of our solutions to spectroscopic parameters.

He-rich atmospheres with temperatures below $\approx$ 12\,000\,K have larger uncertainties on their stellar parameters because of the strong effect of trace hydrogen or metal opacities, leading to differences of up to 0.2\,dex between the pure-He and mixed solutions (see also \citealt{bergeronetal19-1}). We emphasise that while we favour the fixed H/He = 10$^{-5}$ abundance solution for accurate $T_{\rm eff}$ and $\log g$ on average \citep{mccleery20}, it is still not clear if this represents the true H/He abundance or if hydrogen is instead a proxy for missing physics in the models, including metal opacities. For cool He-rich atmospheres with detectable metals, carbon or hydrogen (DZ, DZA, DQ, etc), our parameters should be considered as indicative, as in those cases it is possible to calculate tailored models more appropriate than those used in our catalogue \citep{hollandsetal18-1,coutu19,blouin19}. Nevertheless, \citet{mccleery20} have shown that reasonable parameters can be obtained for all He-rich atmospheres, including DQ and DZ stars, using mixed H/He models above 7000\,K and pure-He models below that temperature, with the resulting mean mass essentially the same as that of DA white dwarfs in the same temperature range. 

\begin{table*}
\centering
\caption{\label{table_format} Our catalogue of white dwarfs includes all columns available in the \textit{Gaia} EDR3 archive. Additional columns unique of this catalogue or not available in the main \textit{Gaia}\,EDR3 distribution are summarized here. The full catalogue will be made available via the VizieR catalogue access tool.}
\begin{tabular}{lll}
\hline
\hline
Column & Heading & Description\\
\hline
1 & \textsc{White\_dwarf\_name} & WD\,J + J2000 ra (hh mm ss.ss) + dec (dd mm ss.s), equinox and epoch 2000\\
3 & \textsc{dr2\_source\_id}& Unique identifier for this object in \textit{Gaia}\,DR2\\
12 & \textsc{ZP\_correction} & Zero point offset correction \citep{lindegrenetal20-1}\footnotemark[3]\\
13 & \textsc{\Pwd} & The \textit{probability of being a white dwarf} (see Sect.\,\ref{main_cut})\\
14 & \textsc{Density} & The number of Gaia sources in the same $\simeq50$ arcsec$^2$ bin as this object (see Sect.\,\ref{main_cut})\\
78 & \textsc{phot\_g\_mean\_flux\_corrected} & Corrected \textsc{phot\_g\_mean\_flux} \citep{Gaia_summary20-1}\footnotemark[2]\\
79 & \textsc{phot\_g\_mean\_mag\_corrected} & Corrected \textsc{phot\_g\_mean\_mag}
\citep{Gaia_summary20-1}\footnotemark[2]\\
80 & \textsc{phot\_g\_mean\_mag\_error\_corrected} & Corrected \textsc{phot\_g\_mean\_mag\_error}
\citep{Gaia_summary20-1}\footnotemark[2]\\
99 & \textsc{phot\_bp\_rp\_excess\_factor\_corrected} & \textsc{phot\_bp\_rp\_excess\_factor} corrected for colour dependence \citep{rielloetal20-1}\footnotemark[1]\\
113 & \textsc{excess\_flux\_error} & Metric for source photometric variability (Sect.\,\ref{var_sect})\\
114 & \textsc{bright\_neighbour} & If 1 it indicates the presence of a source 5 mag brighter than the target in the $G$-band within $5''$\\
115 & \textsc{AV\_mean} & Mean extinction value [mag] derived from 3D reddening maps (Sect.\,\ref{extinction})\\
116 & \textsc{AV\_min}& Extinction value [mag] derived from 3D reddening maps using $-1\sigma$ EDR3 distance (Sect.\,\ref{extinction})\\
117 & \textsc{AV\_max}& Extinction value [mag] derived from 3D reddening maps using $+1\sigma$ EDR3 distance (Sect.\,\ref{extinction})\\
118 & \textsc{flag\_ext} & 0 indicates that the object is located within the 3D extinction map or that real\\ & & extinction value could be < 0.2\, mag larger than \textsc{AV\_max}\\
&   & 1 indicates that the object is located outside the 3D extinction map and that real \\ & & extinction value could be  between 0.2 and 0.5\,mag larger than \textsc{AV\_max} \\
&   & 2 indicates that the object is located outside the 3D extinction map and that real\\ & & extinction value could be  between 0.5 and 1.0\,mag larger than \textsc{AV\_max} \\
&   & 3 indicates that the object is located outside the 3D extinction map and that real\\ & & extinction value could be  >1.0\,mag larger than \textsc{AV\_max} (Sect.\,\ref{extinction})\\
119 & \textsc{\Teff\_(H)} & Effective temperature [K] from fitting the dereddened $G$, $G_{\rm BP}$, and $G_{\rm RP}$ absolute fluxes\\
& & with pure-H model atmospheres (see Sect.\,\ref{sect_params})\\
120 & \textsc{$\sigma$\_\Teff\_(H)} & Uncertainty on \Teff\_(H)~[K]\\
121 & \textsc{$\log$\_$g$\_(H)}& Surface gravity [cm/s$^2$] from fitting the dereddened $G$, $G_{\rm BP}$, and $G_{\rm RP}$ absolute fluxes\\
& & with pure-H model atmospheres (see Sect.\,\ref{sect_params})\\
122 & \textsc{$\sigma$\_$\log\_g$\_(H)} & Uncertainty on $\log\_g$\_(H) [cm/s$^2$]\\
123 & \textsc{$M$\_(WD, H)}& Stellar mass [$\mathrm{M}_{\odot}$] resulting from the adopted mass-radius relation\\
& & and best fit parameters (see Sect.\,\ref{sect_params})\\
124 & \textsc{$\sigma$\_$M$\_(WD, H)} & Uncertainty on \textsc{$M$\_(WD, H)} [$\mathrm{M}_{\odot}$]\\
125 & \textsc{$\chi^2$\_(H)} & $\chi^2$ value of the fit (pure-H)\\
126 & \textsc{\Teff\_(He)} & Effective temperature [K] from fitting the dereddened $G$, $G_{\rm BP}$, and $G_{\rm RP}$ absolute fluxes\\
& & with pure-He model atmospheres (see Sect.\,\ref{sect_params})\\
127 & \textsc{$\sigma$\_\Teff\_(He)} & Uncertainty on \Teff\_(He) [K]\\
128 & \textsc{$\log$\_$g$\_(He)}& Surface gravity [cm/s$^2$] from fitting the dereddened $G$, $G_{\rm BP}$, and $G_{\rm RP}$ absolute fluxes\\
& & with pure-He model atmospheres (see Sect.\,\ref{sect_params})\\
129 & \textsc{$\sigma$\_$\log\_g$\_(He)} & Uncertainty on $\log\_g\_(He)$ [cm/s$^2$]\\
130 & \textsc{$M$\_(WD, He)}& Stellar mass [$\mathrm{M}_{\odot}$] resulting from the adopted mass-radius relation\\
& & and best fit parameters (see Sect.\,\ref{sect_params})\\
131 & \textsc{$\sigma$\_$M$\_(WD, He)} & Uncertainty on \textsc{$M$\_(WD, He)} [$\mathrm{M}_{\odot}$]\\
132 & \textsc{$\chi^2$\_(He)} & $\chi^2$ value of the fit (pure-He)\\
133 & \textsc{\Teff\_(mixed)} & Effective temperature [K] from fitting the dereddened $G$, $G_{\rm BP}$, and $G_{\rm RP}$ absolute fluxes\\
& & with mixed H-He model atmospheres (see Sect.\,\ref{sect_params})\\
134 & \textsc{$\sigma$\_\Teff\_(mixed)} & Uncertainty on \Teff [K]\\
135 & \textsc{$\log$\_$g$\_(mixed)}& Surface gravity [cm/s$^2$] from fitting the dereddened $G$, $G_{\rm BP}$, and $G_{\rm RP}$ absolute fluxes\\
& & with mixed H-He model atmospheres (see Sect.\,\ref{sect_params})\\
136 & \textsc{$\sigma$\_$\log\_g$\_(mixed)} & Uncertainty on $\log g$ [cm/s$^2$]\\
137 & \textsc{$M$\_(WD, mixed)}& Stellar mass [$\mathrm{M}_{\odot}$] resulting from the adopted mass-radius relation\\
& & and best fit parameters (see Sect.\,\ref{sect_params})\\
138 & \textsc{$\sigma$\_$M$\_(WD, mixed)} & Uncertainty on the mass [$\mathrm{M}_{\odot}$]\\
139 & \textsc{$\chi^2$\_(mixed)} & $\chi^2$ value of the fit (mixed H-He)\\
140 & \textsc{r\_med\_geo}& Median of the geometric distance posterior [pc] \citep{bailer-jones21-1}\\
141 & \textsc{r\_lo\_geo}&16th percentile of the geometric distance posterior [pc] \citep{bailer-jones21-1}\\
142 & \textsc{r\_hi\_geo}&84th percentile of the geometric distance posterior [pc] \citep{bailer-jones21-1}\\
143 & \textsc{r\_med\_photpgeo}&Median of the photogeometric distance posterior [pc] \citep{bailer-jones21-1}\\
144 & \textsc{r\_lo\_photogeo}&16th percentile of the photogeometric distance posterior [pc] \citep{bailer-jones21-1}\\
145 & \textsc{r\_hi\_photogeo}&84th percentile of the photogeometric distance posterior [pc] \citep{bailer-jones21-1}\\
146 & \textsc{fidelity\_v1}& `astrometric fidelity' metric from \citet{rybizkietal21-1}\\
147 & \textsc{SDSS\_name} & SDSS object name if available (SDSS + J2000 coordinates) \\
148 & \textsc{SDSS\_clean} & If 1 the SDSS photometry for this object is considered clean\\ 
& &(see https://www.sdss.org/dr16/tutorials/flags)\\

\hline
\end{tabular}
\end{table*}

\addtocounter{table}{-1}
\begin{table*}
\caption{continued from previous page}
\begin{tabular}{lll}
\hline
\hline
Column & Heading & Description\\
\hline
149 & \textsc{$u$mag} & SDSS $u$ band magnitude [mag]\\
150 & \textsc{$u$mag\_err} & SDSS $u$ band magnitude uncertainty [mag]\\
151 & \textsc{$g$mag} & SDSS $g$ band magnitude [mag]\\
152 & \textsc{$g$mag\_err} & SDSS $g$ band magnitude uncertainty [mag]\\
153 & \textsc{$r$mag} & SDSS $r$ band magnitude [mag]\\
154 & \textsc{$r$mag\_err} & SDSS $r$ band magnitude uncertainty [mag]\\
155 & \textsc{$i$mag} & SDSS $i$ band magnitude [mag]\\
156 & \textsc{$i$mag\_err} & SDSS $i$ band magnitude uncertainty [mag]\\
157 & \textsc{$z$mag} & SDSS $z$ band magnitude [mag]\\
158 & \textsc{$z$mag\_err} & SDSS $z$ band magnitude uncertainty [mag]\\
159 & \textsc{SDSS\_separation}& Angular separation between the \textit{Gaia} source and its associated  SDSS object, after coordinate separation [arcsec]\\ 
160 & \textsc{SDSS\_spectra}& Number of SDSS spectra available for this object\\
\hline
\end{tabular}
\end{table*}

\section{The \texorpdfstring{\textit{Gaia}}{Gaia}--SDSS spectroscopic sample}
\label{SDSS_section}
We cross-matched our Gaia catalogue of white dwarf candidates with the SDSS DR16 spectroscopic catalogue and retrieved  38\,740 spectra corresponding to 29\,254 objects. After visual inspection of  the spectra we identified 25\,176 spectroscopically confirmed white dwarfs (with a total of 33\,473 spectra). 473 additional white dwarfs with SDSS spectra (with a total of 483 spectra) were found in the RPM-extension (Sect.\,\ref{sect_ext}). 
Although this spectroscopic subset only covers $\simeq7$~per cent of the over 359\,000 white dwarf candidates in our full catalogue, it still represents the largest sample of spectroscopically confirmed \textit{Gaia} white dwarfs to date, a record that will most likely be kept until new multi-object spectroscopic facilities (WEAVE, DESI, 4MOST, SDSS\,V) will begin to systematically observe white dwarfs. 
In our classification of SDSS spectra we adopted 25
 classes for white dwarfs: DA, DB, DBA, DAB, DO, DAO, DC, DAZ, DZA, DBZ, DZB, DBAZ, DABZ, DZBA, DZAB, DZ, DQ, hotDQ, DQpec, DAH, DBH, DZH, MWD, PG1159, WD (Fig.\,\ref{spec_bars}; see \citealt{sionetal83-1,koester13-1} for the definition of these classes). Objects classified as “MWD” are magnetic white dwarfs where the distortion of spectral features due to the magnetic field is so severe that we were unable to reliably identify the atmospheric composition. Spectra marked as “WD” have spectra too poor for detailed classification in a sub-class, but still broadly recognizable as those of white dwarfs.
 White dwarfs in binaries and non-white dwarf contaminants were grouped in six additional spectral classes (CV, DB+MS, DA+MS, DC+MS, STAR, QSO). Finally, spectra with signal-to-noise ratio too low for visual classification were simply classed as "unreliable". 
 Combined with the \textit{Gaia} EDR3 data and our stellar parameters (Sect.\,\ref{sect_params}), this spectroscopic sample represents an ideal tool to further explore the  global properties of white dwarfs.
 The full \textit{Gaia}-SDSS spectroscopic sample can be downloaded separately from our main catalogue of white dwarfs (Table\,\ref{SDSS_Table}) at the following link:
 \url{https://warwick.ac.uk//fac/sci/physics/research/astro/research/catalogues/gaiaedr3_wd_rpm_ext.fits.gz} 
\begin{figure}
    \centering
    \includegraphics[width=\columnwidth]{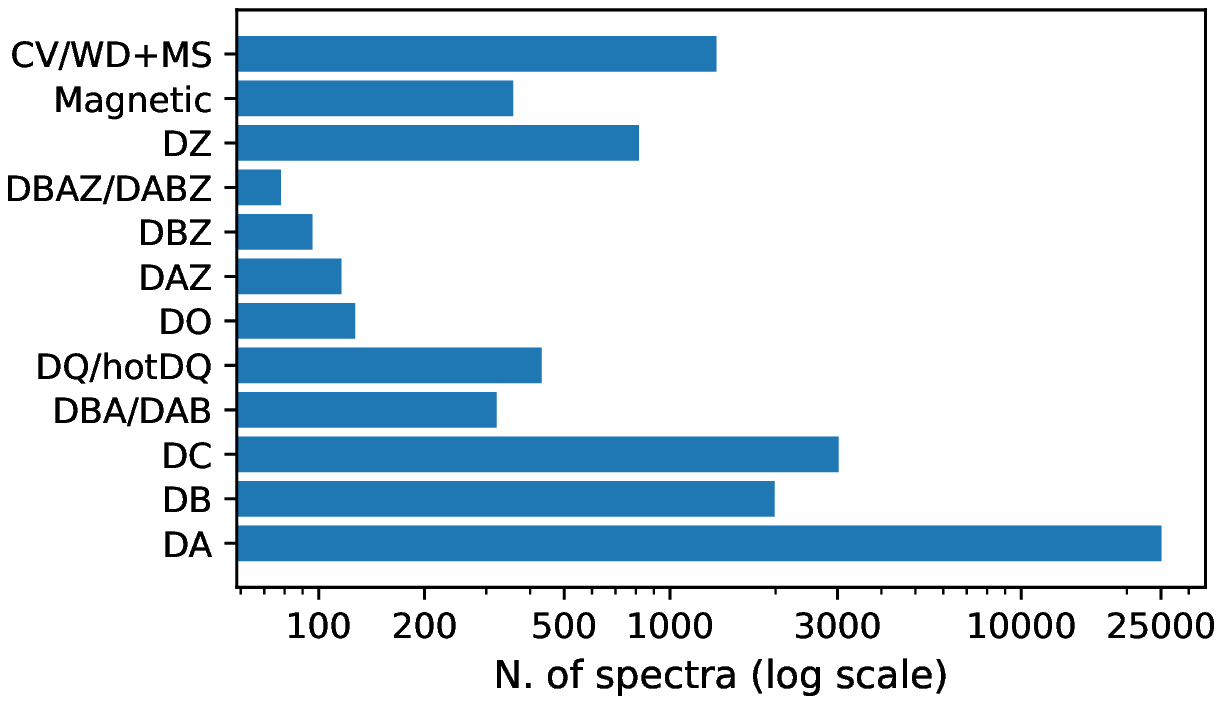}
    \caption{Number of SDSS spectra in each of the main white dwarf spectral classes. }
    \label{spec_bars}
\end{figure}

\begin{table*}
\centering
\caption{\label{SDSS_Table} Description of the columns unique to the \textit{Gaia} EDR3-SDSS spectroscopic catalogue. The full list also includes a number of  columns replicated from our main catalogue and so described in Table\,\ref{table_format} or in the official EDR3 archive distribution. The full catalogue will be made available online via the VizieR catalogue access tool.}
\begin{tabular}{lll}
\hline
\hline
Column No. & Heading & Description\\
  	\hline
4 & \textsc{WD\_catalogue} & M if the object is included in the main catalogue, E if the object is in the RPM extension (Sect.\,\ref{sect_ext})\\
51 & \textsc{Plate} & Identifier of the plate used in the observation of the spectrum\\
52 & \textsc{MJD} & Modified Julian date of the observation of the spectrum \\
53 & \textsc{FiberID} & Identifier of the fiber used in the observation of the spectrum\\
54 & \textsc{S/N} & Signal-to-noise ratio of the spectrum calculated in the range 4500--5500\,\AA\\
55 & \textsc{spec\_class} & Classification of the object based on a visual inspection of the SDSS spectrum\\
\hline
\end{tabular}
\end{table*}
\section{An indicator of intrinsic variability: \texorpdfstring{\textsc{excess\_flux\_error}}{excess\_flux\_error}}
\label{var_sect}
All photometric measurements provided in the \textit{Gaia} archive are produced by combining the multiple observations the spacecraft obtained for each object.  Every individual observation naturally results in slightly different measurements due to a combination of factors including instrumental errors and potentially intrinsic brightness changes in the observed objects. Therefore, the final photometric errors provided in the archive should reflect the overall scatter in the individual \textit{Gaia} measurements of each objects and so depend on the magnitude of the target, its colour and the number of observations. Consequently, intrinsically variable stars should acquire additional error due to the increased scatter in brightness measured by \textit{Gaia} across the different observations. Therefore unusually high photometric errors could be a sign of stellar variability, but to evaluate whether any specific target is a high-error outlier, one has to first establish the typical photometric errors for all objects in the same parameter range (color, flux and number of observation) as the target of interest. 

\begin{figure}
\includegraphics[width=0.95\columnwidth]{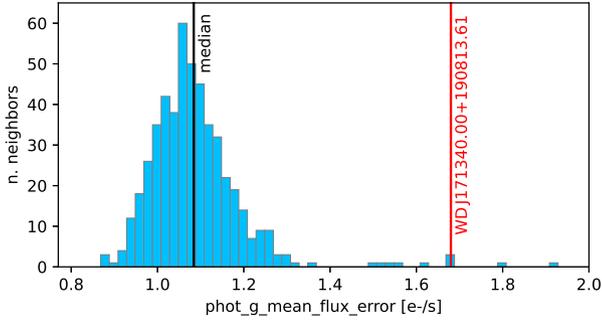}
\caption{\label{var_hist} Distribution of \textsc{phot\_g\_mean\_mag\_error} values of the 500 neighbours of the white dwarf candidate WD\,J171340.00+190813.61 with \textsc{excess\_flux\_error} $\simeq8.6$. The black line indicates the median \textsc{phot\_g\_mean\_flux\_error} of the neighbours and the red line indicates the \textsc{phot\_g\_mean\_flux\_error} of the white dwarf candidate.}
\end{figure}

For each object in our white dwarf catalogue we retrieved the 500 \textit{Gaia} sources which clear the selection criteria in Eqs.\,4 and 8, and are closest in terms \textsc{phot\_g\_mean\_flux}, number of observations in $G$ (\textsc{phot\_g\_n\_obs}) and colour (\textsc{bp\_rp}). 
This was done by taking the 500 closest objects (neighbours) in terms of the Euclidean distance metric in 3-dimensional space. To avoid the dominance of one parameter, they were pre-scaled using a Min-Max approach, i.e the range of values spanned by each parameter was normalised on a linear scale from 0 to 1. We find that blue sources in EDR3 have systematically larger relative photometric errors than redder sources prompting the need to limit the neighbour selection to objects of similar colour. 
The search was also restricted to a specific area of the sky: objects with $|b|<15$ are compared only to other objects within $|b|<15$, and similarly for $|b|\geq15$. This restriction in position takes care of higher errors due to crowding in regions close to the Galactic plane, while selecting only the closest neighbours in the aforementioned 3D parameter space ensures the comparison sample only contains objects with similar instrumental error. In order to systematically compare the photometric errors associated with our target to those of the 500 neighbours, we define the quantity \textsc{excess\_flux\_error} as the ratio of the $\log_{10}$ of an object’s flux error to the median absolute deviation ($MAD$) of the $\log_{10}$ of the flux error of its neighbours:
\begin{multline} \tag{46}
\textsc{excess\_flux\_error} =  \\ \frac 
{\splitfrac{\log_{10}(\textsc{phot\_g\_mean\_flux\_error})}{-\mathrm{median}( \log_{10}(\textsc{phot\_g\_mean\_flux\_error}_\mathrm{neighbours} ))}}
{\mathrm{MAD}(\log_{10}(\textsc{phot\_g\_mean\_flux\_error}_\mathrm{neighbours} ))}
\end{multline}
We calculated \textsc{excess\_flux\_error} for all objects in our catalogue and include it as a column. 
Negative values were set to zero.
High values of \textsc{excess\_flux\_error} (e.g. $>4$) indicate that the target object’s flux error is significantly higher than that of its neighbours (Fig. \ref{var_hist}).

For example, of the 218  Cataclysmic Variables (CV) in our \textit{Gaia}-SDSS spectroscopic sample, 171 have \textsc{excess\_flux\_error} $>4$, indicating that strongly variable sources can be identified by this parameter. Indeed CVs display some the highest \textsc{excess\_flux\_error} in the catalogue with a median value of $\simeq13$.

\begin{figure}
\includegraphics[width=0.95\columnwidth]{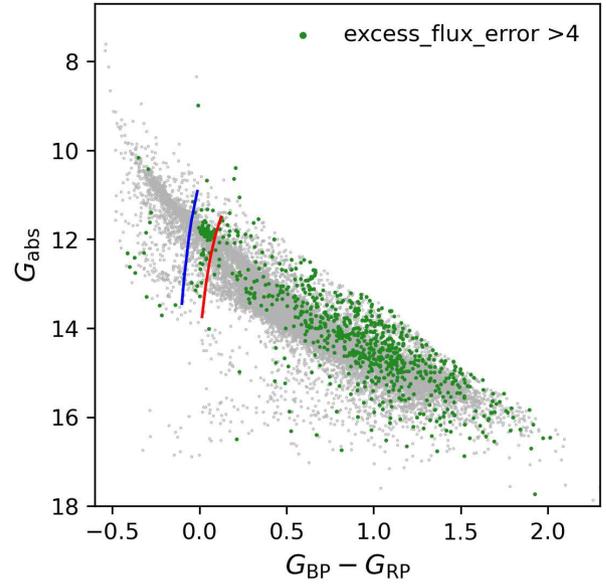}
\caption{\label{var_scatter} Distribution of all objects in our catalogue with $P_{\rm wd} >0.75$ and parallax $\geq$ 10 (grey points). Sources with \textsc{excess\_flux\_error} $>4$ are plotted as green points. The blue and red edges of the  ZZ\,Ceti empirical instability strip \citep{gianninasetal15-1} are indicated by the blue and red line respectively. }
\end{figure}

As a further test of the potential of the \textsc{excess\_flux\_error} as an indicator of intrinsic variability, we used the empirical ZZ\,Ceti instability strip defined in \citet{gianninasetal15-1} 
and reliable stellar parameters for H-atmosphere white dwarfs ($\chi^{2}_{(\mathrm{H})} <1.5$, see Sect.\,\ref{sect_params}) to select 3295 relatively bright pulsating white dwarf candidates with $G<19$. We find that 5.7~per cent of these stars have \textsc{excess\_flux\_error} $>4$, while outside the empirical instability strip only 1.3~per cent of similarly bright white dwarf candidates have \textsc{excess\_flux\_error} $>4$ (Fig.\,\ref{var_scatter}). This simple test indicates that  \textsc{excess\_flux\_error} is, in some capacity, sensitive to the brightness variation caused by white dwarf pulsation. However, only a relative small fraction of objects with this level of variability (amplitude 1--30~per cent, \citealt{mukadametal13-1}) can be reliably identified using this metric. 

\citet{guidryetal20-1} carried out a similar exploration of white dwarf variability in \textit{Gaia}\,DR2 by calculating a variability index with a similar scope to our \textsc{excess\_flux\_error}. 
Though the two metrics are based on different data-sets, if they are both capable of identifying genuine variable \textit{Gaia}  sources, there should be a significant overlap in the samples of variable candidates selected using them. 
We find that about half of the top one~per cent variable white dwarf candidates within 200\,pc selected by \citet{guidryetal20-1} also have  \textsc{excess\_flux\_error} $>4$. Additionally all eight new ZZ Ceti which \citet{guidryetal20-1} identified using their \textit{Gaia} variability index have \textsc{excess\_flux\_error} $>4.6$.

\section{Discussion}
\label{discussion}
\subsection{Overall completeness: Comparison with a SDSS sample of white dwarf candidates}
Some intrinsic limits in \textit{Gaia} observations (e.g. problems with crowding, uneven scanning law, very broad filters) as well as the complex selection method behind our final white dwarf sample may cause some genuine white dwarfs to be excluded from our catalogue. In order to quantitatively estimate the combined impact of these factors, we need to compare the new \textit{Gaia} catalogue of white dwarfs with an independent, sufficiently large and well-defined sample of stellar remnants. 
All the spectroscopic samples of white dwarfs currently available, including SDSS, are severely incomplete and biased by the specific observing strategy adopted, and so ill-suited for this task.
In \citet{gentilefusilloetal19-1} we used an independently constructed sample of white dwarf candidates selected on the basis of their colour and reduced proper motion as described in \citet{gentilefusilloetal15-1}. This sample used SDSS photometry and proper motions from the \textit{Gaia}-PS1-SDSS (GPS1)  \citep{tianetal17-1} catalogue with no additional \textit{Gaia}-based input.
We opt again to use this sample of white dwarf candidates as a comparison group to test the completeness of our new EDR3 white dwarf catalogue. This also allows a direct comparison with the values obtained for our DR2-based catalogue.  
We note that because of the colour restrictions used in the construction of the SDSS comparison sample, it only contains white dwarfs with \Teff\ $>7000$\,K, and an additional $\simeq 14\,000$ are potentially missing because they lacked reliable proper motion measurement in GPS1. We estimated the SDSS comparison sample to contain $\simeq 75$~per cent of all the white dwarfs observed by SDSS, brighter than $g=20.1$ and with \Teff\ $>7000$\,K. For completeness, we also point out that the footprint of the SDSS photometry is mostly limited to high Galactic latitudes with $|b|\gtrsim20^{\circ}$. For a detailed description of the development and characterisation of the SDSS comparison sample, see Appendix A in \citet{gentilefusilloetal19-1}.

Analogously to the procedure in \citet{gentilefusilloetal19-1}, we begin by selecting a subset of 60\,739 high-confidence  dwarf candidates from the SDSS comparison sample. We estimate that this subset only has seven~per cent contamination while still including 97~per cent of all the white dwarfs in the full sample.
We then cross-matched the sky position of these objects (correcting for different epoch of observation) with our \textit{Gaia} EDR3 catalogue of white dwarf candidates and retrieved a total of 52\,465 stars. Accounting for the estimated level of contamination of the SDSS sample, we can use the number of stars not retrieved in our cross-match to estimate an upper limit in the completeness of the \textit{Gaia} EDR3 catalogue of 93 ~per cent. Similarly, we can use the estimated completeness of the SDSS comparison sample and the number of objects retrieved in  cross-match to calculate a lower limit in completeness of 67~per cent. In comparison, when performing the same test for our \textit{Gaia} DR2 catalogue of white dwarfs we estimated a maximum completeness of 85~per cent and a minimum one of 60~per cent. We emphasise that these values can be considered fully descriptive only for white dwarfs with $G$ $\leq20$ and \Teff\ $> 7000$\,K, at high Galactic latitudes ($|b|>20^{\circ}$). 


\begin{figure}
\includegraphics[width=0.95\columnwidth]{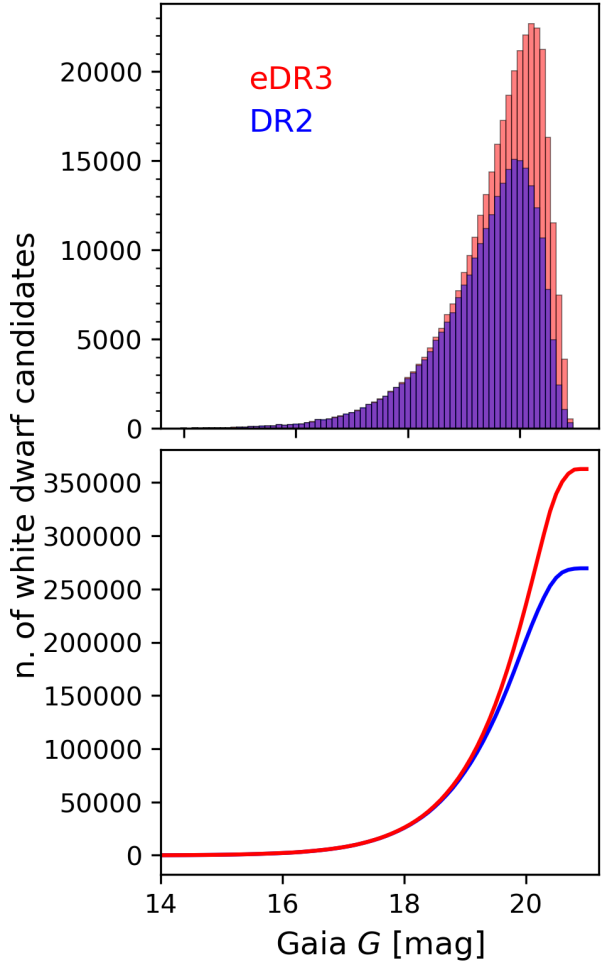}
\caption{\label{G_hist} \textit{Top panel:} Number high-confidence white dwarf candidates from \citet[$\Pwd > 0.75$, blue, DR2]{gentilefusilloetal19-1} and from the catalogue presented here ($\Pwd > 0.75$, red, EDR3)  as function of \textit{Gaia} $G$ magnitude. 
\textit{Bottom panel}: Same as top, but showing the cumulative distribution.}
\end{figure}

\subsection{New sky density and limiting magnitude: Comparison with DR2 catalogue of white dwarf candidates} 
In order to assess the improvements brought on by EDR3 compared to DR2, we compared the newly constructed catalogue described here, to our \textit{Gaia} DR2 catalogue of white dwarfs \citep{gentilefusilloetal19-1}. We cross-matched the two catalogues by directly comparing the unique \textit{Gaia} source IDs using the auxiliary table \textsc{gaiaedr3.dr2\_neighbourhood} provided in the \textit{Gaia} archive. We find that $247\,505$ high-confidence  white dwarfs we selected in DR2 (94~per cent) are again identified as white dwarf candidates with \Pwd\ $>0.75$ in our new EDR3 catalogue. In contrast 5249 objects previously identified as reliable white dwarf candidates, while still included in the EDR3 catalogue, now have \Pwd\ $<0.75$. Additionally 9726 DR2 white dwarf candidates are entirely excluded in the new catalogue. EDR3 photometric and astrometric measurements for these missing sources are 
either considered unreliable according to our selection criteria (Eqs\.\,1-21), or place these objects  in areas of the H-R diagram not occupied by white dwarfs.

\begin{figure}
\includegraphics[width=0.95\columnwidth]{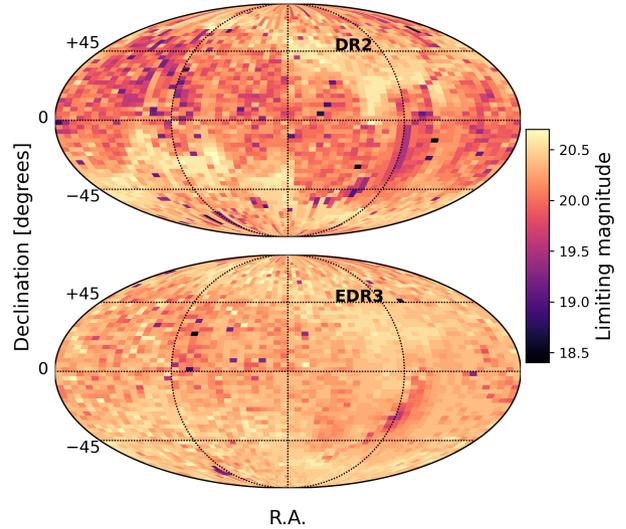}
\caption{\label{Gaia_limit} Limiting magnitude for \textit{Gaia} DR2 white dwarf candidates from \citet[\textit{top panel}]{gentilefusilloetal19-1} and for white dwarf candidates from the new EDR3 catalogue presented here (\textit{bottom panel}). Both are calculated using 10 deg$^{2}$ bins.}
\end{figure}

Our EDR3 catalogue  also includes $\simeq 99\,000$ new white dwarf candidates. As illustrated in Fig.\,\ref{G_hist} the vast majority of the new white dwarfs are fainter than magnitude $G=19$ and for stars brighter than this limit the EDR3 white dwarf catalogue is only eight~per cent larger than its predecessor. This is a direct consequence of the improved depth of the EDR3 observations. Most bright white dwarfs already had robust photometric and astrometric data in DR2, but in EDR3 the limiting magnitude for objects with reliable measurements is significantly more uniform across the entire celestial sphere compared to DR2 (Fig.\,\ref{Gaia_limit}). Virtually all parts of the sky are now covered to a magnitude depth of at least $G=20$, and we estimate the  sky density of white dwarfs with $G \leq 20$ in EDR3 to range from $4.7~\mathrm{deg}^{-2}$ at $|b|>80$ to $6.1~\mathrm{deg}^{-2}$ at $|b|<10$, with an all-sky average of $\simeq 5.6~\mathrm{deg}^{-2}$.

\begin{figure}
\includegraphics[width=0.95\columnwidth]{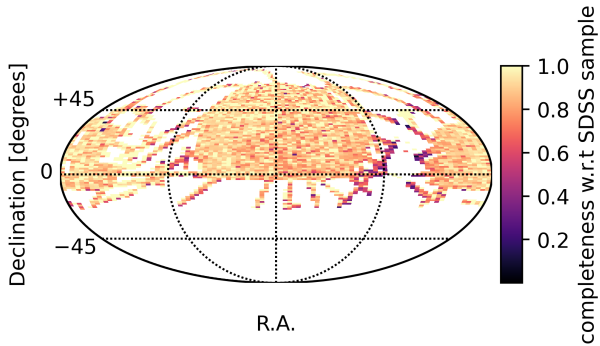}
\caption{\label{sky_hist} Completeness of our \textit{Gaia} catalogue of white dwarf candidates with respect to the SDSS comparison sample, as a function of sky position. Each bin represents 5 deg$^{2}$.}
\end{figure}

The increase in number of newly identified white dwarfs is particularly marked in crowded areas of the sky where the improved EDR3 measurements allows us to lift some of the strict limitations we imposed in DR2 for our selection of white dwarfs in these regions.  Nonetheless, even in EDR3, our ability to  identify white dwarfs from \textit{Gaia} data is still reduced in highly crowded areas of the sky compared to less populated areas and, as a result, the completeness of our catalogue drops in the central regions of the Galactic plane (Fig.\,\ref{sky_hist}).

\begin{table}
\centering
\caption{\label{summary_diff} Summary of the differences in number of selected white dwarf candidates in the \textit{Gaia} EDR3 catalogue presented     here and the DR2 catalogue from \citet{gentilefusilloetal19-1}.}
\begin{tabular}{lr}
\hline\\[-1.5ex]
White dwarf candidates in EDR3 (\Pwd\ $>0.75$) & 359\,073\\
White dwarf candidates in DR2 (\Pwd\ $>0.75$) & 262\,480\\
New candidates in EDR3 & 99\,151\\
\hspace{0.5cm}of which with $G\leq19$ & 2937\\
\hspace{0.5cm}of which completely absent in DR2 & 197\\
\hspace{0.5cm}of which with unreliable or incomplete data in DR2 & 98\,954\\
DR2 candidates not included in EDR3 catalogue & 9726\\
    \hspace{0.5cm}of which in 40pc (according to DR2 parallax)  & 22\\
\hline
\end{tabular}
\end{table}

\subsection{The \texorpdfstring{100\,pc}{100 pc} sample and a comparison with white dwarfs in the Gaia catalogue of nearby stars} 
\begin{figure*}
\includegraphics[width=1.8\columnwidth]{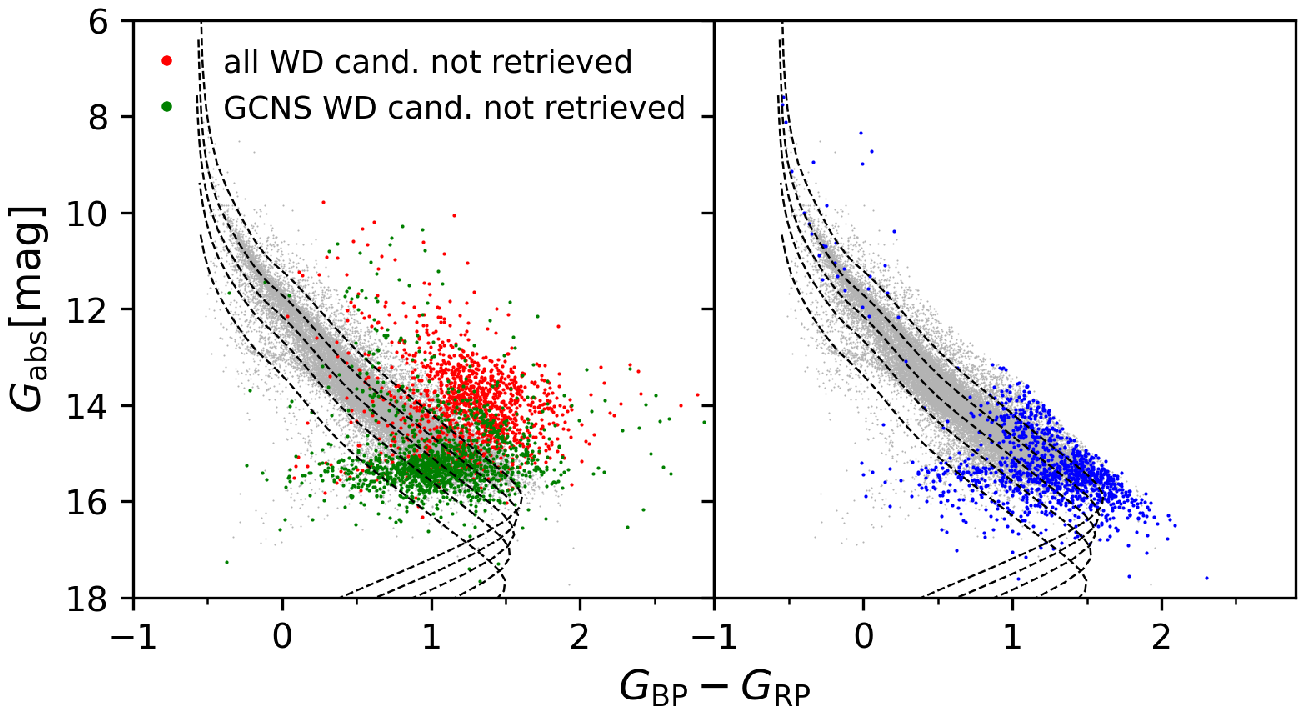}
\caption{\label{gcns_diff} Distribution of objects in common between our catalogue of \textit{Gaia} EDR3 white dwarf candidates (\Pwd\ > 0.75) and the white dwarfs selected by \citet[grey points]{Gaia_GCNS2020-1}. Objects identified as white dwarf candidates in \citet{Gaia_GCNS2020-1} but not recovered by our selection are shown in red, and green if also included in the GCNS (\textit{left panel}). Vice-versa objects identified as nearby white dwarf candidates in this work, but not in \citet{Gaia_GCNS2020-1} are shown in blue  (\textit{right panel}). The black dashed lines indicate the cooling tracks for H-atmosphere white dwarfs with masses between 0.4\,\Msun\ and 1.2\,\Msun\ in steps of 0.2\,\Msun.}
\end{figure*}

In the \textit{Gaia} EDR3 release paper \citet{Gaia_GCNS2020-1} the authors include a catalogue of nearby white dwarfs. These stars were identified using a random forest classifier trained on 20\,000 previously known white dwarfs retrieved from catalogues based on \textit{Gaia DR2} astrometry \citep{gentilefusilloetal19-1, torresetal19-1, jimenez-esteban18-1}, with the only initial constraint of parallax $>8$. The classifier tested 1\,050\,614 EDR3 sources and assigned to each a probability of being a white dwarf. The authors then considered each object with a probability greater than 0.5 as a valid white dwarf and selected a sample of 32\,948 white dwarf candidates. Of these objects 21\,848 also clear all selection criteria to be included in the Gaia Catalogue of Nearby Stars \citep[GCNS]{Gaia_GCNS2020-1}.
From our catalogue we selected a similar sample of high confidence white dwarf candidates (\Pwd\ $>0.75$ and parallax $>8$) and compared it to the selection in \citet{Gaia_GCNS2020-1}. We find the two samples in close agreement with 30\,255 objects in common, but there are also some differences worth of notice. Our catalogue includes 1149 white dwarf candidates not found in \citet{Gaia_GCNS2020-1} and similarly \citet{Gaia_GCNS2020-1} includes 2693 objects not retrieved  by our selection. Even when considering only the 21\,848 white dwarf candidates in the GCNS, 1419 sources are still not present in our sample (Fig.\,\ref{gcns_diff}).  About half the sources in both unmatched samples (not recovered in our catalogue, but selected in the GCNS and vice-versa) are located in dense areas of the Galactic plane where \textit{Gaia} observations are less reliable. Additionally the objects in both these samples form an horizontal cluster in H-R space which does not reflect the predicted locus of white dwarfs, suggesting these objects may not be real white dwarfs (though this effect is more marked for the objects excluded by our catalogue). 
Fig.\,\ref{gcns_diff} also shows two limitations which \citet{Gaia_GCNS2020-1} already identified in their white dwarf selection. Firstly, their sample includes a number of objects located above and red-ward of the white dwarf locus. Most of these stars are not included in the GCNS and \citet{Gaia_GCNS2020-1} observed that they may not be true white dwarfs. We suggest that these objects are likely white dwarf plus M-dwarf binaries and, though a number of them are also included in our catalogue, most have $\Pwd\ <0.75$ which can distinguish them from well-behaved single white dwarfs. Secondly, $\simeq30$ bright and relatively hot white dwarfs (some of them historically well-known, e.g. Sirius B) are missing from the \citet{Gaia_GCNS2020-1} white dwarf sample as these rare objects were not sufficiently represented in the training sample used by the random forest algorithm.
From this comparison it is apparent that selecting the coolest and reddest white dwarfs, even within $\simeq 100$\,pc, remains a challenging endeavour, and contamination of the red section of the white dwarf cooling sequence from other sources is something users should be aware of even in EDR3. Consequently it is not possibly to conclusively state weather the \textit{Gaia} EDR3 100\,pc white dwarf sample can be considered complete.

\subsection{Comparison with `astrometric fidelity' \texorpdfstring{\citep{rybizkietal21-1}}{(Rybiziki et al. 2021)}}
Recently \citet{rybizkietal21-1} presented a novel approach to 
quantify the robustness of the astrometric solution for sources in \textit{Gaia} EDR3. Their method relies on a neural net that uses 14 pertinent \textit{Gaia} parameters to capture the overall reliability of the \textit{Gaia} measurements  in a single `astrometric fidelity’ parameter. 

\begin{figure}
\includegraphics[width=0.95\columnwidth]{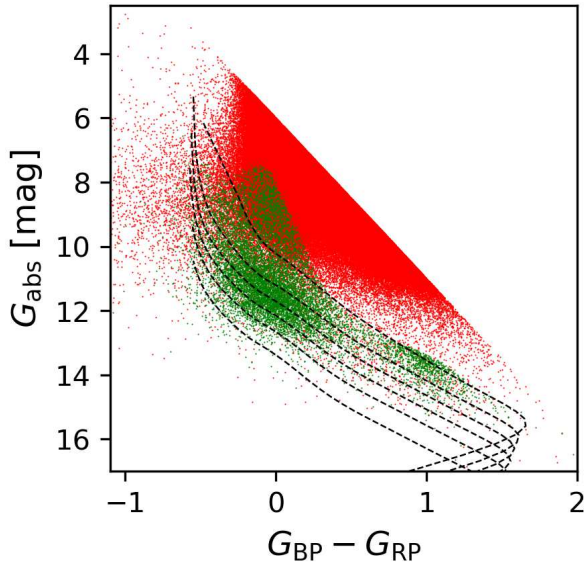}
\caption{\label{fidel_compare}  Distribution of "high-fidelity sources" from \citep[][`astrometric fidelity' > 0.5]{rybizkietal21-1} brighter than $G=20$ excluded from our catalogue (red points), and of "low-fidelity sources" (`astrometric fidelity' < 0.5) which are instead included with \Pwd~> 0.75 (green points). The black dashed lines indicate the cooling tracks for H-atmosphere white dwarfs with masses between 0.4\,\Msun\ and 1.2\,\Msun\ in steps of 0.2\,\Msun.}
\end{figure}

We retrieved this `astrometric fidelity’ for all sources within our initial selection (Eqs.\,1-2) and used it as an independent test to verify the robustness of our quality filtering (Eqs.\,3-21).
We find that our cuts in \textsc{astrometric\_sigma5d\_max}, \textsc{parallax\_over\_error} and \textsc{pm\_over\_err} eliminate the vast majority of objects with very low `astrometric fidelity' ($<0.1$), confirming the strength of these criteria as discriminators of spurious sources. However our final sample excludes $\simeq 50$~per cent of the "high-fidelity sources" (`astrometric fidelity'$>0.5$) while approximately half of the objects included are "low-fidelity sources" (`astrometric fidelity'$<0.5$). The majority of the "high-fidelity sources" excluded are located in the densest areas of the Galactic plane or in the Magellanic clouds and are rejected by our cut on \textsc{corrected\_excess\_factor}. Additionally 
most of these rejected objects sit on the boundary of our cut in H-R space (i.e. not on the white dwarf sequence, Fig.\,\ref{fidel_compare}).

On the other hand we also find that $\simeq40\,000$ objects in our catalogue with $\Pwd>0.75$ are "low-fidelity sources" (11\,411 with $G <$ 20). A large fraction of these sources cluster in the upper portion of the white dwarf cooling sequence (Fig.\,\ref{fidel_compare}) and are most likely moderately hot white dwarfs. Additionally 1243 SDSS spectroscopically confirmed white dwarfs  are among these "low-fidelity sources" which we identified as reliable high-confidence white dwarf candidates.

In conclusion, the comparison reveals a marked discrepancy between our quality selection and a selection done relying exclusively on the `astrometric fidelity' computed by \citet{rybizkietal21-1}. This is not surprising as, in the current version of their algorithm \citet{rybizkietal21-1} do not use \textsc{phot\_bp\_rp\_excess\_factor} as discriminating parameter and their training sample is constructed from sources with a match in the Two Micron All Sky Survey (2MASS) and so it is not well suited for faint blue sources like white dwarfs.
Nonetheless, the `astrometric fidelity' could still be useful to a number of users and we decided to include it in our catalogue under the column \textsc{fidelity\_v1}.

\subsection{Volume complete samples}

The local volume complete sample of white dwarfs, with the Sun at its center, has historically increased from a radius of 13\,pc \citep{holbergetal02-1} to 20--40\,pc \citep{giammicheleetal12-1,sionetal14-1,limoges15}, and finally to $\approx$ 15\,000 white dwarf candidates within 100\,pc from \textit{Gaia} DR2 and EDR3 \citep{jimenez-esteban18,gentilefusilloetal19-1,Gaia_GCNS2020-1}. However the vast majority of white dwarf candidates in that \textit{Gaia} identified sample lack follow-up spectroscopy. Unbiased spectroscopic samples are critically important as they are benchmarks to characterise local space density, stellar formation history, binary evolution or white dwarf spectral evolution and crystallisation. Currently the largest volume complete samples with near-complete ($>98$~per cent) spectroscopic confirmation are the overlapping 20\,pc \citep{hollands18} and northern 40\,pc volumes \citep{mccleery20} as identified from \textit{Gaia} DR2. Work is also progressing towards spectroscopic follow-up of the southern 40\,pc sample (in preparation) and 100\,pc SDSS footprint \citep{kilic20}.

In the following subsections, we describe how our new EDR3 white dwarf catalogue has changed the completeness and properties of the existing 20\,pc and 40\,pc volume samples as described specifically by our earlier publications \citep{hollands18,mccleery20,gentilefusilloetal19-1}. 

\subsubsection{\texorpdfstring{20\,pc}{20 pc} volume sample}

Compared to both \citet{hollands18} and \citet{gentilefusilloetal19-1}, the 20\,pc volume sample has five new sources in the EDR3 catalogue. Four of them are well known white dwarfs with missing or incomplete complete data in DR2 (40\,Eri\,B, Ross\,627, EGGR\,290 and Wolf\,489). Another known white dwarf (WD\,0728+642) moved from just outside 20\,pc to just inside the volume, although it is still within 1$\sigma$ of the boundary. 
No confirmed 20\,pc DR2 white dwarf is missing in our new catalogue or has moved out of the sample. However, four confirmed 20\,pc white dwarfs (Procyon~B, WD\,0208$-$510, WD\,0727+482A and WD\,0727+482B) are missing or have incomplete data in both DR2 and EDR3.

Using the EDR3 catalogue, we have determined a revised estimate of the local white dwarf space-density.
We repeated the procedure outlined in \citet{hollands18}, where the main difference is the four previously missing objects
from DR2 now present in EDR3 as described above. This acts to increase the volume-averaged detection efficiency to
$98.3_{-1.2}^{+0.6}$~per cent (compared with $96.0_{-1.6}^{+1.3}$~per cent for DR2), corresponding to an effective
volume sampled by \emph{Gaia} of $32\,950_{-390}^{+230}$\,pc$^3$ (from $32\,170_{-540}^{+420}$\,pc$^3$ for DR2).
This increase in effective volume is compensated by for increased number of objects detected in EDR3, thus
leading to a revised space-density of $4.47\pm0.37\times10^{-3}$\,pc$^{-3}$~--~almost unchanged from our DR2 estimate of $4.49\pm0.38\times10^{-3}$\,pc$^{-3}$.
The approach of \citet{hollands18}, also yields the distribution for the number of white dwarfs which remain
undiscovered within 20\,pc. As with DR2, the most likely value is zero, but with its probability increased from
28~per cent to 37~per cent. The median of this distribution has decreased from two one, i.e. there is only a 40~per cent
probability that more than one white dwarf remains undetected within the 20\,pc volume.

\subsubsection{\texorpdfstring{40\,pc}{40 pc} volume sample}
\label{40pc}

There are 33  new sources in our catalogue and within 40\,pc that were not in \citet{gentilefusilloetal19-1} and within the same volume. Nine are known white dwarfs that had incomplete or missing data in DR2 such that they could not be reliably identified as white dwarfs, while five are known white dwarfs that have moved from $>$40\,pc to $<$40\,pc. All 14 known white dwarfs have \Pwd\ $>$ 0.75. The faintest confirmed white dwarf in the 40\,pc sample is $G = 19.61$, and we note that only two new candidates are fainter.

The majority of the 19 remaining new sources are either wide companions to bright known main-sequence stars (with separation from 5 to 20 arcsec), or are near bright background sources, or are in crowded fields. We suspect that \textit{Gaia} colours may be unreliable for sources close to bright companions of magnitude $G \approx 6-12$. Consequently we warn users that some of these new white dwarf candidates may potentially be  misidentified low-mass stars with poor background subtraction or contamination from a companion, although a few could also be genuine new Sirius-like systems.

As a consequence, it is difficult to estimate the number of genuine new EDR3 white dwarfs within 40\,pc, but it could be only a handful. To give the catalogue users some indication of which objects may be affected by the presence of a bright neighbour we provide a \textsc{bright\_neighbour} flag. This flag is given to all stars with a neighbour $\simeq5$ magnitude brighter then themselves at a separation of $5 \arcsec$ or less. A number of extremely bright stars are not included at all in \textit{Gaia} EDR3, so to compute our \textsc{bright\_neighbour} flag we also performed a cross match with the Tycho-2 catalogue of bright stars \citep{hogetal00-1}.

No confirmed 40\,pc white dwarf identified in DR2 is missing from our new catalogue, but three confirmed white dwarfs have moved from within 40\,pc to outside of the volume: WD\,J102459.83+044610.50, 
WD\,J065722.88+024100.84 and 
WD\,J214810.74$-$562613.14. There are 18 additional high probability candidates from DR2 that are gone. None are known to be white dwarfs but 10 were confirmed as main-sequence stars in \citet{tremblay20}. Finally, 162 low probability white dwarf candidates from \citet{gentilefusilloetal19-1} are now entirely gone from the 40\,pc sample. None of them had been identified as genuine white dwarf by spectroscopic follow-up. 

Now we focus our attention on the northern 40\,pc \textit{Gaia} sample with already high spectroscopic completeness, with all membership changes listed in Table~\ref{tab:appendix_40pc}. The full EDR3 northern hemisphere 40\,pc sample can also be queried via the \href{ https://vizier.u-strasbg.fr/viz-bin/VizieR?-source=J/MNRAS/499/1890}{VizieR catalogue access tool} as a direct update of the tables presented in \citet{mccleery20}.

Of the 521 spectroscopically confirmed \textit{Gaia} DR2 white dwarfs from \citet{mccleery20}, two have now moved beyond 40\,pc as discussed above (category B of Table~\ref{tab:appendix_40pc}). This is compensated by 12 previously known white dwarfs moving in or now having full \textit{Gaia} solutions (category A of Table~\ref{tab:appendix_40pc}), for a new total of 531 spectroscopically confirmed white dwarfs. However, several externally confirmed 40\,pc members are still missing from our EDR3 catalogue (categories C and D of Table~\ref{tab:appendix_40pc}).

Three unobserved high probability DR2 candidates (category E of Table~\ref{tab:appendix_40pc}) are still strong candidates in EDR3 and likely white dwarfs. Furthermore, five unobserved low probability white dwarf candidates in table A2 of \citet{mccleery20} are now high probability candidates in our catalogue (category E of Table~\ref{tab:appendix_40pc}), hence likely white dwarfs. Even accounting for the few new EDR3 candidates that may be genuine white dwarfs, we estimate that only $\simeq10$ white dwarfs within 40\,pc north have not yet received spectroscopic follow-up, corresponding to a spectroscopic completeness of $>$98~per cent. 

\begin{figure}
\includegraphics[width=0.94\columnwidth]{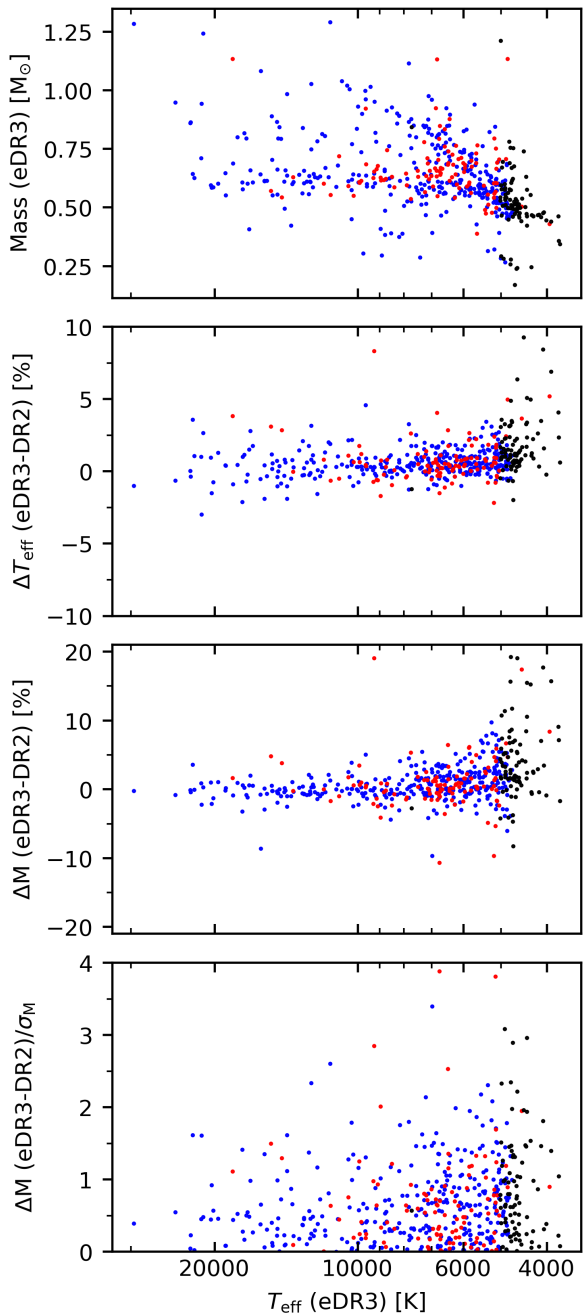}
\caption{\label{pier0} {\it Top panel}: EDR3 $T_{\rm eff}$ versus mass distribution for the northern 40\,pc white dwarf sample. The sample is similar to that of \citet{mccleery20} but with updated EDR3 parameters and minor changes in membership (see Table~\ref{tab:appendix_40pc}). Spectrally identified H-rich atmospheres are shown in blue, He-rich atmospheres in red, and unconstrained compositions in black (cool DC white dwarfs and unobserved candidates). {\it Top middle panel:} Difference in $T_{\rm eff}$ (per cent) between EDR3 and DR2. {\it Bottom middle panel:} Difference in mass (per cent) between EDR3 and DR2. {\it Bottom panel:} Absolute difference in mass divided by EDR3 mass uncertainty.}
\end{figure}

\subsection{Change in white dwarf stellar parameters}
\label{parameters40pc}

To characterise changes in derived white dwarf fundamental parameters between EDR3 and DR2, we rely on the well characterised northern 40\,pc sample. We compare the published DR2 parameters from table A1 of \citet{mccleery20} with the updated EDR3 Vizier version of the same table (see Sect.\,\ref{40pc}). Fig.~\ref{pier0} (top panel) shows the resulting EDR3 $T_{\rm eff}$ versus mass distribution, which is very similar to that found in \citet{mccleery20} using DR2 and the same model atmospheres. To illustrate this we plot EDR3 versus DR2 differences in $T_{\rm eff}$ and mass in the middle panels of Fig.~\ref{pier0}. It demonstrates that despite changes in passbands and colour calibration \citep{rielloetal20-1}, there is almost negligible systematic changes between DR2 and EDR3 parameters. While $T_{\rm eff}$ depends almost only on \textit{Gaia} colours, mass is sensitive to both changes in colours ($T_{\rm eff}$) and parallax. The $\Delta M$ scatter is almost twice as large as the $\Delta T_{\rm eff}$ scatter, therefore suggesting that changes in colours and parallax contribute roughly equally to changes in mass (or $\log g$). Finally, the bottom panel of Fig.~\ref{pier0} compares absolute changes in mass between catalogues and our formal mass error bars which are based solely on catalogued \textit{Gaia} data errors. The median change in mass is 0.57$\sigma$, which demonstrates that despite  \textit{Gaia} error bars being very small, they appear to be a good representation of the precision of the data. Nevertheless, this does not account for possible systematic issues in both DR2 and EDR3 that could impact the accuracy of \textit{Gaia} atmospheric parameters (see Sect.\,\ref{elena}).

We note that reddening is very small and essentially negligible for the 40\,pc sample. This is not the case for our overall EDR3 catalogue, where stellar parameters are also modified by our newly adopted extinction maps.

\subsection{Calibration of Gaia EDR3 colours}
\label{elena}
\citet{cukanovaite2021} tested the calibration of \textit{Gaia} DR2 colours, by comparing
spectroscopically- and photometrically-derived parameters for various samples of DA and DB white dwarfs.
In this section, we reproduce the test performed in \cite{cukanovaite2021}, using our stellar parameters  based on \textit{Gaia} EDR3 photometry (see Sect.\,\ref{sect_params}) and spectroscopically-derived parameters recovered from various other studies. The DA white dwarfs used in this comparison are from two samples: the SDSS sample from \cite{tremblay2019} and the \cite{gianninas2011} sample. The latter sample has been corrected for 3D effects by \citet{tremblay2019}. The DB samples are from \cite{genest2019b} and \citet{rolland2018}, with additional correction for 3D effects from \cite{cukanovaite2021}. The \cite{genest2019b} sample has also been corrected to put van der Waals broadening on the same scale as the other samples. For a more detailed discussion on these corrections, see \cite{cukanovaite2021}. 

We cross-matched the coordinates of all white dwarfs in the spectroscopic samples with our  \textit{Gaia} EDR3 catalogue and recovered the photometrically-derived effective temperatures and surface gravities. Only a small percentage of the white dwarfs did not have a match in our catalogue because of missing or unreliable \textit{Gaia} EDR3 parallaxes or colours. 

For all successful matches, the differences in spectroscopic and photometric effective temperatures and surface gravities were calculated. For the final comparison, we removed all white dwarfs that had absolute differences larger than 30~per cent. This was done to remove any physical outliers, such as unresolved binaries. This clipping affects only a very small percentage of objects, with the exception of the \cite{rolland2018} sample, where at very low effective temperatures the high-$\log{g}$ problem is apparent when the removal is not performed. Nevertheless, this does not change our conclusions. For a full overview of the \cite{rolland2018} sample without the removal of outliers see figure 15 in \cite{cukanovaite2021}. 

\begin{figure}
	\includegraphics[width=0.99\columnwidth]{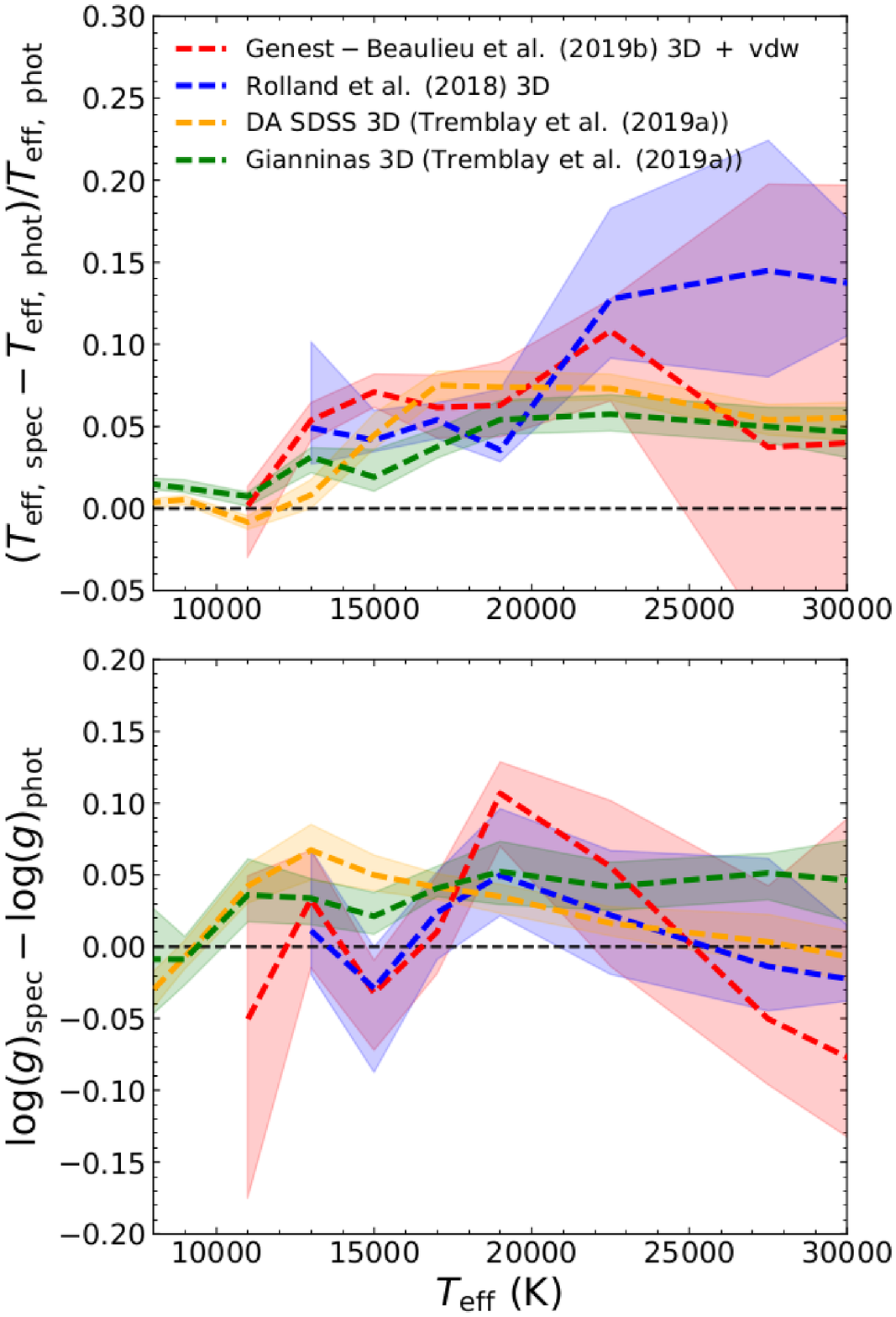}
    \caption{A comparison between EDR3 \Teff\ and $\log g$ values calculated in this paper and 
    corresponding spectroscopically-derived parameters from literature. The spectroscopic parameters for the two DA samples are from \protect\cite{tremblay2019}. The DB spectroscopic parameters are from \protect\cite{genest2019b} and \protect\cite{rolland2018}, and have been corrected for 3D effects by \protect\cite{cukanovaite2021}. In the case of \protect\cite{genest2019b}, the sample has also been corrected for van der Waals effects (see \protect\citealt{cukanovaite2021}). The dashed lines indicate the medians for the various samples. The coloured areas represent the error on the median found by bootstrapping. The median difference was calculated in bins of 2000\,K for $T_{\rm{eff}} < 20\,000$\,K and in bins of 5000 K for $T_{\rm{eff}} > 20\,000$\,K.}
    
    \label{fig:phot_spec_comp}
\end{figure}

Fig.~\ref{fig:phot_spec_comp} displays the results of our comparison of spectroscopically- and photometrically-derived parameters. The plot shows the median difference in bins of 2000 K for $T_{\rm{eff}} < 20\,000$ K, and in bins of 5000 K for $T_{\rm{eff}} > 20\,000$ K. By comparing this figure with figures 14 and 15 in \cite{cukanovaite2021}, it is clear that the offsets seen in the spectroscopically- and photometrically-derived parameters are very similar for  \textit{Gaia} DR2 and EDR3. 

A $\simeq5$~per cent offset in effective temperature can be seen around 20\,000 K. Analogously to what was concluded by \cite{cukanovaite2021}, we attribute this to issues with \textit{Gaia} colour calibrations. 

In particular, we note that the spectrophotometric calibration of the Space Telescope Imaging Spectrograph (STIS) on board the Hubble Space Telescope (HST) is tied to the spectroscopic parameters of three white dwarfs that are also part of the \citet{gianninas2011} sample adopted here. Therefore, the systematic $T_{\rm{eff}}$ offset observed in Fig.~\ref{fig:phot_spec_comp} suggests that \textit{Gaia} EDR3 is still not entirely consistent with \textit{HST}/STIS calibration \citep{maiz18}.

Other explanations could include issues with the calibration of the surveys from which the spectroscopic samples were derived, such as SDSS. However, SDSS and non-SDSS spectroscopic samples show similar offsets. Another explanation could be issues with the microphysics of model spectra, such as prescriptions of line broadening. We believe this to be a less likely explanation because line broadening theories for DA and DB white dwarfs are entirely different and have different temperature dependencies. Thus, it is difficult to see how the offset could be so similar between the two types of white dwarf samples.

\section{Conclusion}
We present a catalogue of white dwarf candidates selected from  \textit{Gaia}\,EDR3.
Starting from the entire 1.47 billion sources with an astrometric solution in EDR3, we defined a number of cuts in colour, absolute magnitude and various \textit{Gaia} quality parameters to  broadly isolate the white dwarf locus 
in the H-R diagram and remove objects with unreliable \textit{Gaia} measurements. This selection resulted in a sample of 1\,280\,266 EDR3 sources. 
Following the same methodology described for DR2 in \citet{gentilefusilloetal19-1} we then made use of a sample SDSS spectroscopically classified objects to map the distribution of white dwarfs and contaminant objects in $\bp-\rp$ colour - \Gabs\ space and calculate  \textit{probabilities of being a white dwarf} (\Pwd) for all objects in our sample. 
Coupled with \textit{Gaia} quality flags these \Pwd\ values allow to flexibly select samples of white dwarf candidates compromising between completeness and contamination according to the users' specific goals.  
For general purpose we recommend a cut at $\Pwd>0.75$, which selects a sample of 359\,073 objects. This subset includes the vast majority of the white dwarfs in the catalogue, with minimal level of contamination. We also utilized state-of-the-art 3D extinction maps to correct the three \textit{Gaia} photometric bands for reddening, and provide stellar parameters (\Teff, $\log g$ and mass) obtained by fitting \textit{Gaia} photometry and parallax for all objects with $\Pwd>0.70$. 
In addition to stellar parameters and \Pwd\ values, our catalogue includes a number of other columns not available in the main \textit{Gaia} EDR3 archival distribution, which can be used to further characterize any  sample of white dwarf candidates selected. For example the parameter \textsc{excess\_flux\_error} is an indicator of the variability in the flux of an object during the multiple \textit{Gaia} observations.  

To assess the overall completeness of the main \textit{Gaia} EDR3 catalogue of white dwarfs, we have used an independent sample of 60\,739 white dwarf candidates selected from SDSS on the basis of their colours and proper motions. We estimate our \textit{Gaia} EDR3 catalogue to be between 67 and 93~per cent complete for white dwarfs with $G$ $\leq20$ and \Teff\ $> 7000$\,K, at high Galactic latitudes ($|b|>20^{\circ}$). 
Together with the main catalogue of white dwarf candidates we also provide an extension containing objects with unreliable parallax measurements (\textsc{parallax\_over\_error} $<1$), but which can be identified as white dwarfs using their reduced proper motion. An additional $\simeq10\,200$ white dwarfs can be recovered from this RPM-extension.

We also cross matched both the main catalogue and the RPM-extension with the entire spectral library of SDSS\,DR16, retrieving a total of 39\,223 spectra corresponding to 29\,727 objects.  All spectra were visually inspected and classified according to their spectral type. We provide this \textit{Gaia}-SDSS spectroscopic catalogue in a separate distribution from the main catalogue of white dwarf candidates.

The catalogue presented in this paper is by far the largest collection of white dwarfs published to date exceeding the size of the largest DR2 based catalogue by nearly 100\,000 objects. With EDR3 virtually every part of the sky now has \textit{Gaia} parameters robust enough to identify all white dwarfs with $G\leq20$. We therefore do not expect significant further improvements in the overall number of known white dwarfs until reliable astrometry for fainter stars will become available thanks to next generation observatories (e.g. Vera Rubin observatory \citealt{LSST19-1}). However, to date only a small fraction of these \textit{Gaia} white dwarfs has received any spectroscopic observations. This type of follow-up is fundamental in order to study white dwarfs in detail, both as individual objects and as a stellar population. 
With new large multi-fibre spectroscopic facilities now beginning operation in both hemispheres (e.g., WEAVE, 4MOST, DESI, SDSS-V, \citealt{weave14-1,4most14-1,DESI16-1,sdssv17-1}), we are on the verge of a revolution in observational astronomy, and spectroscopic coverage of every known white dwarf may become a reality in the not-too-distant future. Our \Pwd\ values combined with stellar parameters provided, and the various \textit{Gaia} metrics, allow to construct well defined white dwarf samples for any spectroscopic follow-up campaign, making the EDR3 catalogue of white dwarf candidates presented here a key resource for any future white dwarf study.

\section*{Acknowledgements}
P.E.T has received funding from the European Research Council under the European Union's Horizon 2020 research and innovation programme n. 677706 (WD3D). R.L. acknowledges access to results from the EXPLORE project prior to publication. EXPLORE has received funding from the European Union’s Horizon 2020 research and innovation programme under grant agreement No 101004214.
This work has made use of data from the European Space Agency (ESA) mission
{\it Gaia} (\url{https://www.cosmos.esa.int/gaia}), processed by the {\it Gaia}
Data Processing and Analysis Consortium (DPAC,
\url{https://www.cosmos.esa.int/web/gaia/dpac/consortium}). Funding for the DPAC
has been provided by national institutions, in particular the institutions
participating in the {\it Gaia} Multilateral Agreement.
Funding for the Sloan Digital Sky 
Survey IV has been provided by the 
Alfred P. Sloan Foundation, the U.S. 
Department of Energy Office of 
Science, and the Participating 
Institutions. 

SDSS-IV acknowledges support and 
resources from the Center for High 
Performance Computing  at the 
University of Utah. The SDSS 
website is www.sdss.org.

SDSS-IV is managed by the 
Astrophysical Research Consortium 
for the Participating Institutions 
of the SDSS Collaboration including 
the Brazilian Participation Group, 
the Carnegie Institution for Science, 
Carnegie Mellon University, Center for 
Astrophysics | Harvard \& 
Smithsonian, the Chilean Participation 
Group, the French Participation Group, 
Instituto de Astrof\'isica de 
Canarias, The Johns Hopkins 
University, Kavli Institute for the 
Physics and Mathematics of the 
Universe (IPMU) / University of 
Tokyo, the Korean Participation Group, 
Lawrence Berkeley National Laboratory, 
Leibniz Institut f\"ur Astrophysik 
Potsdam (AIP),  Max-Planck-Institut 
f\"ur Astronomie (MPIA Heidelberg), 
Max-Planck-Institut f\"ur 
Astrophysik (MPA Garching), 
Max-Planck-Institut f\"ur 
Extraterrestrische Physik (MPE), 
National Astronomical Observatories of 
China, New Mexico State University, 
New York University, University of 
Notre Dame, Observat\'ario 
Nacional / MCTI, The Ohio State 
University, Pennsylvania State 
University, Shanghai 
Astronomical Observatory, United 
Kingdom Participation Group, 
Universidad Nacional Aut\'onoma 
de M\'exico, University of Arizona, 
University of Colorado Boulder, 
University of Oxford, University of 
Portsmouth, University of Utah, 
University of Virginia, University 
of Washington, University of 
Wisconsin, Vanderbilt University, 
and Yale University.

This work was funded by the Deutsche Forschungsgemeinschaft (DFG, German Research Foundation) -- Project-ID 138713538 -- SFB 881 (``The Milky Way System'').

\section*{Data Availability}
The catalogues presented in this work can be downloaded HERE. They will also be made available via the VizieR Service for Astronomical Catalogues. 
All additional data underlying this article is publicly available from the relevant survey archives or will be shared on reasonable request to the corresponding author.


\bibliographystyle{mnras}
\bibliography{aamnem99,aabib1,aabib2,aabib_new,aabib_tremblay,aabib_elena,Rosinebib}

\begin{thebibliography}{}
\makeatletter
\relax
\def\mn@urlcharsother{\let\do\@makeother \do\$\do\&\do\#\do\^\do\_\do\%\do\~}
\def\mn@doi{\begingroup\mn@urlcharsother \@ifnextchar [ {\mn@doi@}
  {\mn@doi@[]}}
\def\mn@doi@[#1]#2{\def\@tempa{#1}\ifx\@tempa\@empty \href
  {http://dx.doi.org/#2} {doi:#2}\else \href {http://dx.doi.org/#2} {#1}\fi
  \endgroup}
\def\mn@eprint#1#2{\mn@eprint@#1:#2::\@nil}
\def\mn@eprint@arXiv#1{\href {http://arxiv.org/abs/#1} {{\tt arXiv:#1}}}
\def\mn@eprint@dblp#1{\href {http://dblp.uni-trier.de/rec/bibtex/#1.xml}
  {dblp:#1}}
\def\mn@eprint@#1:#2:#3:#4\@nil{\def\@tempa {#1}\def\@tempb {#2}\def\@tempc
  {#3}\ifx \@tempc \@empty \let \@tempc \@tempb \let \@tempb \@tempa \fi \ifx
  \@tempb \@empty \def\@tempb {arXiv}\fi \@ifundefined
  {mn@eprint@\@tempb}{\@tempb:\@tempc}{\expandafter \expandafter \csname
  mn@eprint@\@tempb\endcsname \expandafter{\@tempc}}}

\bibitem[\protect\citeauthoryear{{Abril}, {Schmidtobreick}, {Ederoclite}  \&
  {L{\'o}pez-Sanjuan}}{{Abril} et~al.}{2020}]{abriletal20-1}
{Abril} J.,  {Schmidtobreick} L.,  {Ederoclite} A.,   {L{\'o}pez-Sanjuan} C.,
  2020, \mn@doi [\mnras] {10.1093/mnrasl/slz181}, \href
  {https://ui.adsabs.harvard.edu/abs/2020MNRAS.492L..40A} {492, L40}

\bibitem[\protect\citeauthoryear{{Bailer-Jones}, {Rybizki}, {Fouesneau},
  {Demleitner}  \& {Andrae}}{{Bailer-Jones} et~al.}{2021}]{bailer-jones21-1}
{Bailer-Jones} C.~A.~L.,  {Rybizki} J.,  {Fouesneau} M.,  {Demleitner} M.,
  {Andrae} R.,  2021, \mn@doi [\aj] {10.3847/1538-3881/abd806}, \href
  {https://ui.adsabs.harvard.edu/abs/2021AJ....161..147B} {161, 147}

\bibitem[\protect\citeauthoryear{{Bauer}, {Schwab}, {Bildsten}  \&
  {Cheng}}{{Bauer} et~al.}{2020}]{bauer20}
{Bauer} E.~B.,  {Schwab} J.,  {Bildsten} L.,   {Cheng} S.,  2020, \mn@doi
  [\apj] {10.3847/1538-4357/abb5a5}, \href
  {https://ui.adsabs.harvard.edu/abs/2020ApJ...902...93B} {902, 93}

\bibitem[\protect\citeauthoryear{{B{\'e}dard}, {Bergeron}, {Brassard}  \&
  {Fontaine}}{{B{\'e}dard} et~al.}{2020}]{bedardetal20-1}
{B{\'e}dard} A.,  {Bergeron} P.,  {Brassard} P.,   {Fontaine} G.,  2020, arXiv
  e-prints, \href {https://ui.adsabs.harvard.edu/abs/2020arXiv200807469B} {p.
  arXiv:2008.07469}

\bibitem[\protect\citeauthoryear{{Bergeron}}{{Bergeron}}{2001}]{bergeron01-1}
{Bergeron} P.,  2001, ApJ, \href {2001ApJ...558..369B} {558, 369}

\bibitem[\protect\citeauthoryear{{Bergeron} et~al.,}{{Bergeron}
  et~al.}{2011}]{bergeron11}
{Bergeron} P.,  et~al., 2011, \mn@doi [\apj] {10.1088/0004-637X/737/1/28},
  \href {http://adsabs.harvard.edu/abs/2011ApJ...737...28B} {737, 28}

\bibitem[\protect\citeauthoryear{{Bergeron}, {Dufour}, {Fontaine}, {Coutu},
  {Blouin}, {Genest-Beaulieu}, {B{\'e}dard}  \& {Rolland}}{{Bergeron}
  et~al.}{2019}]{bergeronetal19-1}
{Bergeron} P.,  {Dufour} P.,  {Fontaine} G.,  {Coutu} S.,  {Blouin} S.,
  {Genest-Beaulieu} C.,  {B{\'e}dard} A.,   {Rolland} B.,  2019, \mn@doi [\apj]
  {10.3847/1538-4357/ab153a}, \href
  {https://ui.adsabs.harvard.edu/abs/2019ApJ...876...67B} {876, 67}

\bibitem[\protect\citeauthoryear{{Blouin} \& {Dufour}}{{Blouin} \&
  {Dufour}}{2019}]{blouin19}
{Blouin} S.,  {Dufour} P.,  2019, \mn@doi [\mnras] {10.1093/mnras/stz2915},
  \href {https://ui.adsabs.harvard.edu/abs/2019MNRAS.490.4166B} {490, 4166}

\bibitem[\protect\citeauthoryear{{Blouin}, {Daligault}  \& {Saumon}}{{Blouin}
  et~al.}{2021}]{blouin21}
{Blouin} S.,  {Daligault} J.,   {Saumon} D.,  2021, arXiv e-prints, \href
  {https://ui.adsabs.harvard.edu/abs/2021arXiv210312892B} {p. arXiv:2103.12892}

\bibitem[\protect\citeauthoryear{{Bohlin}, {Gordon}  \& {Tremblay}}{{Bohlin}
  et~al.}{2014}]{bohlinetal14-1}
{Bohlin} R.~C.,  {Gordon} K.~D.,   {Tremblay} P.~E.,  2014, \mn@doi [\pasp]
  {10.1086/677655}, \href
  {https://ui.adsabs.harvard.edu/abs/2014PASP..126..711B} {126, 711}

\bibitem[\protect\citeauthoryear{{Camisassa}, {Althaus}, {Torres},
  {C{\'o}rsico}, {Rebassa-Mansergas}, {Tremblay}, {Cheng}  \&
  {Raddi}}{{Camisassa} et~al.}{2021}]{cami21}
{Camisassa} M.~E.,  {Althaus} L.~G.,  {Torres} S.,  {C{\'o}rsico} A.~H.,
  {Rebassa-Mansergas} A.,  {Tremblay} P.-E.,  {Cheng} S.,   {Raddi} R.,  2021,
  \mn@doi [\aap] {10.1051/0004-6361/202140720}, \href
  {https://ui.adsabs.harvard.edu/abs/2021A&A...649L...7C} {649, L7}

\bibitem[\protect\citeauthoryear{{Capitanio}, {Lallement}, {Vergely},
  {Elyajouri}  \& {Monreal-Ibero}}{{Capitanio} et~al.}{2017}]{capitanio17}
{Capitanio} L.,  {Lallement} R.,  {Vergely} J.~L.,  {Elyajouri} M.,
  {Monreal-Ibero} A.,  2017, \mn@doi [\aap] {10.1051/0004-6361/201730831},
  \href {https://ui.adsabs.harvard.edu/abs/2017A&A...606A..65C} {606, A65}

\bibitem[\protect\citeauthoryear{{Chandra}, {Hwang}, {Zakamska}  \&
  {Cheng}}{{Chandra} et~al.}{2020}]{chandraetal20-1}
{Chandra} V.,  {Hwang} H.-C.,  {Zakamska} N.~L.,   {Cheng} S.,  2020, \mn@doi
  [\apj] {10.3847/1538-4357/aba8a2}, \href
  {https://ui.adsabs.harvard.edu/abs/2020ApJ...899..146C} {899, 146}

\bibitem[\protect\citeauthoryear{{Cheng}, {Cummings}  \& {M{\'e}nard}}{{Cheng}
  et~al.}{2019}]{cheng2019}
{Cheng} S.,  {Cummings} J.~D.,   {M{\'e}nard} B.,  2019, \mn@doi [\apj]
  {10.3847/1538-4357/ab4989}, \href
  {https://ui.adsabs.harvard.edu/abs/2019ApJ...886..100C} {886, 100}

\bibitem[\protect\citeauthoryear{{Coutu}, {Dufour}, {Bergeron}, {Blouin},
  {Loranger}, {Allard}  \& {Dunlap}}{{Coutu} et~al.}{2019}]{coutu19}
{Coutu} S.,  {Dufour} P.,  {Bergeron} P.,  {Blouin} S.,  {Loranger} E.,
  {Allard} N.~F.,   {Dunlap} B.~H.,  2019, \mn@doi [\apj]
  {10.3847/1538-4357/ab46b9}, \href
  {https://ui.adsabs.harvard.edu/abs/2019ApJ...885...74C} {885, 74}

\bibitem[\protect\citeauthoryear{{Cukanovaite}, {Tremblay}, {Bergeron},
  {Freytag}, {Ludwig}  \& {Steffen}}{{Cukanovaite}
  et~al.}{2021}]{cukanovaite2021}
{Cukanovaite} E.,  {Tremblay} P.-E.,  {Bergeron} P.,  {Freytag} B.,  {Ludwig}
  H.-G.,   {Steffen} M.,  2021, \mn@doi [\mnras] {10.1093/mnras/staa3684},
  \href {https://ui.adsabs.harvard.edu/abs/2021MNRAS.501.5274C} {501, 5274}

\bibitem[\protect\citeauthoryear{{DESI Collaboration} et~al.,}{{DESI
  Collaboration} et~al.}{2016}]{DESI16-1}
{DESI Collaboration} et~al., 2016, preprint, \href
  {http://adsabs.harvard.edu/abs/2016arXiv161100036D} {} (\mn@eprint {arXiv}
  {1611.00036})

\bibitem[\protect\citeauthoryear{{Dalton} et~al.,}{{Dalton}
  et~al.}{2014}]{weave14-1}
{Dalton} G.,  et~al., 2014, in Society of Photo-Optical Instrumentation
  Engineers (SPIE) Conference Series. p.~0 (\mn@eprint {arXiv} {1412.0843}),
  \mn@doi{10.1117/12.2055132}

\bibitem[\protect\citeauthoryear{{Dennihy} et~al.,}{{Dennihy}
  et~al.}{2020}]{dennihyetal20-2}
{Dennihy} E.,  et~al., 2020, arXiv e-prints, \href
  {https://ui.adsabs.harvard.edu/abs/2020arXiv201003693D} {p. arXiv:2010.03693}

\bibitem[\protect\citeauthoryear{{Dufour}, {Blouin}, {Coutu},
  {Fortin-Archambault}, {Thibeault}, {Bergeron}  \& {Fontaine}}{{Dufour}
  et~al.}{2017}]{dufouretal17-1}
{Dufour} P.,  {Blouin} S.,  {Coutu} S.,  {Fortin-Archambault} M.,  {Thibeault}
  C.,  {Bergeron} P.,   {Fontaine} G.,  2017, in {Tremblay} P.-E.,  {Gaensicke}
  B.,   {Marsh} T.,  eds,  Astronomical Society of the Pacific Conference
  Series Vol. 509, 20th European White Dwarf Workshop. p.~3 (\mn@eprint {arXiv}
  {1610.00986})

\bibitem[\protect\citeauthoryear{{Eisenstein} et~al.,}{{Eisenstein}
  et~al.}{2006}]{eisensteinetal06-1}
{Eisenstein} D.~J.,  et~al., 2006, \mn@doi [ApJS] {10.1086/507110}, 167, 40

\bibitem[\protect\citeauthoryear{{El-Badry}, {Rix}  \& {Weisz}}{{El-Badry}
  et~al.}{2018}]{el-badryetal18-1}
{El-Badry} K.,  {Rix} H.-W.,   {Weisz} D.~R.,  2018, \mn@doi [ApJ Lett.]
  {10.3847/2041-8213/aaca9c}, \href
  {http://adsabs.harvard.edu/abs/2018ApJ...860L..17E} {860, L17}

\bibitem[\protect\citeauthoryear{{Fabricius} et~al.,}{{Fabricius}
  et~al.}{2020}]{fabriciusetal20-1}
{Fabricius} C.,  et~al., 2020, arXiv e-prints, \href
  {https://ui.adsabs.harvard.edu/abs/2020arXiv201206242F} {p. arXiv:2012.06242}

\bibitem[\protect\citeauthoryear{{Fontaine}, {Brassard}  \&
  {Bergeron}}{{Fontaine} et~al.}{2001}]{fontaineetal01-1}
{Fontaine} G.,  {Brassard} P.,   {Bergeron} P.,  2001, PASP, \href
  {2001PASP..113..409F} {113, 409}

\bibitem[\protect\citeauthoryear{{Gaia Collaboration}, {Brown}, {Vallenari},
  {Prusti}, {de Bruijne}, {Babusiaux}  \& {Biermann}}{{Gaia Collaboration}
  et~al.}{2020a}]{Gaia_summary20-1}
{Gaia Collaboration} {Brown} A.~G.~A.,  {Vallenari} A.,  {Prusti} T.,  {de
  Bruijne} J.~H.~J.,  {Babusiaux} C.,   {Biermann} M.,  2020a, arXiv e-prints,
  \href {https://ui.adsabs.harvard.edu/abs/2020arXiv201201533G} {p.
  arXiv:2012.01533}

\bibitem[\protect\citeauthoryear{{Gaia Collaboration} et~al.,}{{Gaia
  Collaboration} et~al.}{2020b}]{gaia-collaboration20-1}
{Gaia Collaboration} et~al., 2020b, arXiv e-prints, \href
  {https://ui.adsabs.harvard.edu/abs/2020arXiv201201771G} {p. arXiv:2012.01771}

\bibitem[\protect\citeauthoryear{{Gaia Collaboration} et~al.,}{{Gaia
  Collaboration} et~al.}{2020c}]{Gaia_GCNS2020-1}
{Gaia Collaboration} et~al., 2020c, arXiv e-prints, \href
  {https://ui.adsabs.harvard.edu/abs/2020arXiv201202061G} {p. arXiv:2012.02061}

\bibitem[\protect\citeauthoryear{{G{\"a}nsicke}, {Schreiber}, {Toloza},
  {Fusillo}, {Koester}  \& {Manser}}{{G{\"a}nsicke}
  et~al.}{2019}]{gaensickeetal19-1}
{G{\"a}nsicke} B.~T.,  {Schreiber} M.~R.,  {Toloza} O.,  {Fusillo} N. P.~G.,
  {Koester} D.,   {Manser} C.~J.,  2019, \mn@doi [\nat]
  {10.1038/s41586-019-1789-8}, \href
  {https://ui.adsabs.harvard.edu/abs/2019Natur.576...61G} {576, 61}

\bibitem[\protect\citeauthoryear{{Genest-Beaulieu} \&
  {Bergeron}}{{Genest-Beaulieu} \& {Bergeron}}{2019a}]{genest19a}
{Genest-Beaulieu} C.,  {Bergeron} P.,  2019a, \mn@doi [\apj]
  {10.3847/1538-4357/aafac6}, \href
  {https://ui.adsabs.harvard.edu/abs/2019ApJ...871..169G} {871, 169}

\bibitem[\protect\citeauthoryear{{Genest-Beaulieu} \&
  {Bergeron}}{{Genest-Beaulieu} \& {Bergeron}}{2019b}]{genest2019b}
{Genest-Beaulieu} C.,  {Bergeron} P.,  2019b, \mn@doi [\apj]
  {10.3847/1538-4357/ab379e}, \href
  {https://ui.adsabs.harvard.edu/abs/2019ApJ...882..106G} {882, 106}

\bibitem[\protect\citeauthoryear{Gentile~Fusillo, G{\"a}nsicke  \&
  Greiss}{Gentile~Fusillo et~al.}{2015}]{gentilefusilloetal15-1}
Gentile~Fusillo N.~P.,  G{\"a}nsicke B.~T.,   Greiss S.,  2015, \mn@doi [MNRAS]
  {10.1093/mnras/stv120}, 448, 2260

\bibitem[\protect\citeauthoryear{{Gentile Fusillo} et~al.,}{{Gentile Fusillo}
  et~al.}{2019}]{gentilefusilloetal19-1}
{Gentile Fusillo} N.~P.,  et~al., 2019, \mn@doi [\mnras]
  {10.1093/mnras/sty3016}, \href
  {https://ui.adsabs.harvard.edu/abs/2019MNRAS.482.4570G} {482, 4570}

\bibitem[\protect\citeauthoryear{{Gentile Fusillo}, {Tremblay}, {Bohlin},
  {Deustua}  \& {Kalirai}}{{Gentile Fusillo}
  et~al.}{2020}]{gentilefusilloetal20-1}
{Gentile Fusillo} N.~P.,  {Tremblay} P.-E.,  {Bohlin} R.~C.,  {Deustua} S.~E.,
   {Kalirai} J.~S.,  2020, \mn@doi [\mnras] {10.1093/mnras/stz2984}, \href
  {https://ui.adsabs.harvard.edu/abs/2020MNRAS.491.3613G} {491, 3613}

\bibitem[\protect\citeauthoryear{{Gentile Fusillo} et~al.,}{{Gentile Fusillo}
  et~al.}{2021}]{gentilefusilloetal21-1}
{Gentile Fusillo} N.~P.,  et~al., 2021, \mn@doi [\mnras]
  {10.1093/mnras/stab992}, \href
  {https://ui.adsabs.harvard.edu/abs/2021MNRAS.tmp..967G} {}

\bibitem[\protect\citeauthoryear{{Giammichele}, {Bergeron}  \&
  {Dufour}}{{Giammichele} et~al.}{2012}]{giammicheleetal12-1}
{Giammichele} N.,  {Bergeron} P.,   {Dufour} P.,  2012, \mn@doi [ApJ]
  {10.1088/0067-0049/199/2/29}, \href
  {http://adsabs.harvard.edu/abs/2012ApJS..199...29G} {199, 29}

\bibitem[\protect\citeauthoryear{{Gianninas}, {Bergeron}  \&
  {Ruiz}}{{Gianninas} et~al.}{2011}]{gianninas2011}
{Gianninas} A.,  {Bergeron} P.,   {Ruiz} M.~T.,  2011, \mn@doi [\apj]
  {10.1088/0004-637X/743/2/138}, \href
  {https://ui.adsabs.harvard.edu/abs/2011ApJ...743..138G} {743, 138}

\bibitem[\protect\citeauthoryear{{Gianninas}, {Kilic}, {Brown}, {Canton}  \&
  {Kenyon}}{{Gianninas} et~al.}{2015}]{gianninasetal15-1}
{Gianninas} A.,  {Kilic} M.,  {Brown} W.~R.,  {Canton} P.,   {Kenyon} S.~J.,
  2015, \mn@doi [\apj] {10.1088/0004-637X/812/2/167}, \href
  {https://ui.adsabs.harvard.edu/abs/2015ApJ...812..167G} {812, 167}

\bibitem[\protect\citeauthoryear{{Guidry} et~al.,}{{Guidry}
  et~al.}{2020}]{guidryetal20-1}
{Guidry} J.~A.,  et~al., 2020, arXiv e-prints, \href
  {https://ui.adsabs.harvard.edu/abs/2020arXiv201200035G} {p. arXiv:2012.00035}

\bibitem[\protect\citeauthoryear{{Hansen}}{{Hansen}}{1998}]{hansen98}
{Hansen} B.~M.~S.,  1998, \mn@doi [\nat] {10.1038/29710}, \href
  {http://adsabs.harvard.edu/abs/1998Natur.394..860H} {394, 860}

\bibitem[\protect\citeauthoryear{{Harris} et~al.,}{{Harris}
  et~al.}{2006}]{harrisetal06-1}
{Harris} H.~C.,  et~al., 2006, \mn@doi [\aj] {10.1086/497966}, \href
  {https://ui.adsabs.harvard.edu/abs/2006AJ....131..571H} {131, 571}

\bibitem[\protect\citeauthoryear{{Hirsch} et~al.,}{{Hirsch}
  et~al.}{2019}]{hirsch19}
{Hirsch} L.~A.,  et~al., 2019, \mn@doi [\apj] {10.3847/1538-4357/ab1b11}, \href
  {https://ui.adsabs.harvard.edu/abs/2019ApJ...878...50H} {878, 50}

\bibitem[\protect\citeauthoryear{{H{\o}g} et~al.,}{{H{\o}g}
  et~al.}{2000}]{hogetal00-1}
{H{\o}g} E.,  et~al., 2000, A\&A, \href {2000A&A...355L..27H} {355, L27}

\bibitem[\protect\citeauthoryear{{Holberg}, {Oswalt}  \& {Sion}}{{Holberg}
  et~al.}{2002}]{holbergetal02-1}
{Holberg} J.~B.,  {Oswalt} T.~D.,   {Sion} E.~M.,  2002, \mn@doi [ApJ]
  {10.1086/339842}, \href {http://adsabs.harvard.edu/abs/2002ApJ...571..512H}
  {571, 512}

\bibitem[\protect\citeauthoryear{{Hollands}, {G{\"a}nsicke}  \&
  {Koester}}{{Hollands} et~al.}{2018a}]{hollandsetal18-1}
{Hollands} M.~A.,  {G{\"a}nsicke} B.~T.,   {Koester} D.,  2018a, \mn@doi
  [\mnras] {10.1093/mnras/sty592}, \href
  {https://ui.adsabs.harvard.edu/abs/2018MNRAS.477...93H} {477, 93}

\bibitem[\protect\citeauthoryear{{Hollands}, {Tremblay}, {G{\"a}nsicke},
  {Gentile-Fusillo}  \& {Toonen}}{{Hollands} et~al.}{2018b}]{hollands18}
{Hollands} M.~A.,  {Tremblay} P.~E.,  {G{\"a}nsicke} B.~T.,  {Gentile-Fusillo}
  N.~P.,   {Toonen} S.,  2018b, \mn@doi [\mnras] {10.1093/mnras/sty2057}, \href
  {https://ui.adsabs.harvard.edu/abs/2018MNRAS.480.3942H} {480, 3942}

\bibitem[\protect\citeauthoryear{{Hollands} et~al.,}{{Hollands}
  et~al.}{2020}]{hollandsetal20-1}
{Hollands} M.~A.,  et~al., 2020, \mn@doi [Nature Astronomy]
  {10.1038/s41550-020-1028-0}, \href
  {https://ui.adsabs.harvard.edu/abs/2020NatAs...4..663H} {4, 663}

\bibitem[\protect\citeauthoryear{{Hollands}, {Tremblay}, {G{\"a}nsicke},
  {Koester}  \& {Gentile-Fusillo}}{{Hollands} et~al.}{2021}]{hollandsetal21-1}
{Hollands} M.~A.,  {Tremblay} P.-E.,  {G{\"a}nsicke} B.~T.,  {Koester} D.,
  {Gentile-Fusillo} N.~P.,  2021, \mn@doi [Nature Astronomy]
  {10.1038/s41550-020-01296-7}, \href
  {https://ui.adsabs.harvard.edu/abs/2021NatAs.tmp...25H} {}

\bibitem[\protect\citeauthoryear{{Inight}, {G{\"a}nsicke}, {Breedt}, {Marsh},
  {Pala}  \& {Raddi}}{{Inight} et~al.}{2021}]{inightetal21-1}
{Inight} K.,  {G{\"a}nsicke} B.~T.,  {Breedt} E.,  {Marsh} T.~R.,  {Pala}
  A.~F.,   {Raddi} R.,  2021, \mn@doi [\mnras] {10.1093/mnras/stab753}, \href
  {https://ui.adsabs.harvard.edu/abs/2021MNRAS.504.2420I} {504, 2420}

\bibitem[\protect\citeauthoryear{{Ivanova}, {Lallement}, {Vergely}  \&
  {Hottier}}{{Ivanova} et~al.}{2021}]{Ivanova21}
{Ivanova} A.,  {Lallement} R.,  {Vergely} J.~L.,   {Hottier} C.,  2021, arXiv
  e-prints, \href {https://ui.adsabs.harvard.edu/abs/2021arXiv210414227I} {p.
  arXiv:2104.14227}

\bibitem[\protect\citeauthoryear{{Ivezi{\'c}} et~al.,}{{Ivezi{\'c}}
  et~al.}{2019}]{LSST19-1}
{Ivezi{\'c}} {\v{Z}}.,  et~al., 2019, \mn@doi [\apj]
  {10.3847/1538-4357/ab042c}, \href
  {https://ui.adsabs.harvard.edu/abs/2019ApJ...873..111I} {873, 111}

\bibitem[\protect\citeauthoryear{{Jim{\'e}nez-Esteban}, {Torres},
  {Rebassa-Mansergas}, {Skorobogatov}, {Solano}, {Cantero}  \&
  {Rodrigo}}{{Jim{\'e}nez-Esteban} et~al.}{2018a}]{jimenez-esteban18}
{Jim{\'e}nez-Esteban} F.~M.,  {Torres} S.,  {Rebassa-Mansergas} A.,
  {Skorobogatov} G.,  {Solano} E.,  {Cantero} C.,   {Rodrigo} C.,  2018a,
  \mn@doi [\mnras] {10.1093/mnras/sty2120}, \href
  {https://ui.adsabs.harvard.edu/abs/2018MNRAS.480.4505J} {480, 4505}

\bibitem[\protect\citeauthoryear{{Jim{\'e}nez-Esteban}, {Torres},
  {Rebassa-Mansergas}, {Skorobogatov}, {Solano}, {Cantero}  \&
  {Rodrigo}}{{Jim{\'e}nez-Esteban} et~al.}{2018b}]{jimenez-esteban18-1}
{Jim{\'e}nez-Esteban} F.~M.,  {Torres} S.,  {Rebassa-Mansergas} A.,
  {Skorobogatov} G.,  {Solano} E.,  {Cantero} C.,   {Rodrigo} C.,  2018b,
  \mn@doi [\mnras] {10.1093/mnras/sty2120}, \href
  {https://ui.adsabs.harvard.edu/abs/2018MNRAS.480.4505J} {480, 4505}

\bibitem[\protect\citeauthoryear{{Jones}}{{Jones}}{1972}]{jones72-1}
{Jones} E.~M.,  1972, \mn@doi [\apj] {10.1086/151701}, \href
  {https://ui.adsabs.harvard.edu/abs/1972ApJ...177..245J} {177, 245}

\bibitem[\protect\citeauthoryear{{Kaiser}, {Clemens}, {Blouin}, {Dufour},
  {Hegedus}, {Reding}  \& {B{\'e}dard}}{{Kaiser} et~al.}{2021}]{kaiseretal21-1}
{Kaiser} B.~C.,  {Clemens} J.~C.,  {Blouin} S.,  {Dufour} P.,  {Hegedus} R.~J.,
   {Reding} J.~S.,   {B{\'e}dard} A.,  2021, \mn@doi [Science]
  {10.1126/science.abd1714}, \href
  {https://ui.adsabs.harvard.edu/abs/2021Sci...371..168K} {371, 168}

\bibitem[\protect\citeauthoryear{{Kawka} \& {Vennes}}{{Kawka} \&
  {Vennes}}{2006}]{kawka06}
{Kawka} A.,  {Vennes} S.,  2006, \mn@doi [\apj] {10.1086/501451}, \href
  {https://ui.adsabs.harvard.edu/abs/2006ApJ...643..402K} {643, 402}

\bibitem[\protect\citeauthoryear{{Kawka}, {Vennes}  \& {Ferrario}}{{Kawka}
  et~al.}{2020}]{kawkaetal20-1}
{Kawka} A.,  {Vennes} S.,   {Ferrario} L.,  2020, \mn@doi [\mnras]
  {10.1093/mnrasl/slz165}, \href
  {https://ui.adsabs.harvard.edu/abs/2020MNRAS.491L..40K} {491, L40}

\bibitem[\protect\citeauthoryear{{Kilic} et~al.,}{{Kilic}
  et~al.}{2006}]{kilicetal06-3}
{Kilic} M.,  et~al., 2006, \mn@doi [\aj] {10.1086/497962}, \href
  {https://ui.adsabs.harvard.edu/abs/2006AJ....131..582K} {131, 582}

\bibitem[\protect\citeauthoryear{{Kilic}, {Bergeron}, {Kosakowski}, {Brown},
  {Ag{\"u}eros}  \& {Blouin}}{{Kilic} et~al.}{2020}]{kilic20}
{Kilic} M.,  {Bergeron} P.,  {Kosakowski} A.,  {Brown} W.~R.,  {Ag{\"u}eros}
  M.~A.,   {Blouin} S.,  2020, \mn@doi [\apj] {10.3847/1538-4357/ab9b8d}, \href
  {https://ui.adsabs.harvard.edu/abs/2020ApJ...898...84K} {898, 84}

\bibitem[\protect\citeauthoryear{{Kilic}, {Bergeron}, {Blouin}  \&
  {B{\'e}dard}}{{Kilic} et~al.}{2021}]{kilicetal21-1}
{Kilic} M.,  {Bergeron} P.,  {Blouin} S.,   {B{\'e}dard} A.,  2021, \mn@doi
  [\mnras] {10.1093/mnras/stab767}, \href
  {https://ui.adsabs.harvard.edu/abs/2021MNRAS.tmp..758K} {}

\bibitem[\protect\citeauthoryear{{Koester}}{{Koester}}{2013}]{koester13-1}
{Koester} D.,  2013, {White Dwarf Stars}.
p.~559, \mn@doi{10.1007/978-94-007-5615-1\_11}

\bibitem[\protect\citeauthoryear{{Kollmeier} et~al.,}{{Kollmeier}
  et~al.}{2017}]{sdssv17-1}
{Kollmeier} J.~A.,  et~al., 2017, arxiv:1711.03234

\bibitem[\protect\citeauthoryear{{Kowalski} \& {Saumon}}{{Kowalski} \&
  {Saumon}}{2006}]{kowalski06}
{Kowalski} P.~M.,  {Saumon} D.,  2006, \mn@doi [\apjl] {10.1086/509723}, \href
  {http://adsabs.harvard.edu/abs/2006ApJ...651L.137K} {651, L137}

\bibitem[\protect\citeauthoryear{{Lallement}, {Vergely}, {Valette},
  {Puspitarini}, {Eyer}  \& {Casagrande}}{{Lallement}
  et~al.}{2014}]{lallement14}
{Lallement} R.,  {Vergely} J.~L.,  {Valette} B.,  {Puspitarini} L.,  {Eyer} L.,
    {Casagrande} L.,  2014, \mn@doi [\aap] {10.1051/0004-6361/201322032}, \href
  {https://ui.adsabs.harvard.edu/abs/2014A&A...561A..91L} {561, A91}

\bibitem[\protect\citeauthoryear{{Lallement}, {Babusiaux}, {Vergely}, {Katz},
  {Arenou}, {Valette}, {Hottier}  \& {Capitanio}}{{Lallement}
  et~al.}{2019}]{lallement19}
{Lallement} R.,  {Babusiaux} C.,  {Vergely} J.~L.,  {Katz} D.,  {Arenou} F.,
  {Valette} B.,  {Hottier} C.,   {Capitanio} L.,  2019, \mn@doi [\aap]
  {10.1051/0004-6361/201834695}, \href
  {https://ui.adsabs.harvard.edu/abs/2019A&A...625A.135L} {625, A135}

\bibitem[\protect\citeauthoryear{{Lam} et~al.,}{{Lam}
  et~al.}{2019}]{lametal19-1}
{Lam} M.~C.,  et~al., 2019, \mn@doi [\mnras] {10.1093/mnras/sty2710}, \href
  {https://ui.adsabs.harvard.edu/abs/2019MNRAS.482..715L} {482, 715}

\bibitem[\protect\citeauthoryear{{Leggett} et~al.,}{{Leggett}
  et~al.}{2018}]{leggett18}
{Leggett} S.~K.,  et~al., 2018, \mn@doi [\apjs] {10.3847/1538-4365/aae7ca},
  \href {https://ui.adsabs.harvard.edu/abs/2018ApJS..239...26L} {239, 26}

\bibitem[\protect\citeauthoryear{{Limoges}, {Bergeron}  \&
  {L{\'e}pine}}{{Limoges} et~al.}{2015}]{limoges15}
{Limoges} M.~M.,  {Bergeron} P.,   {L{\'e}pine} S.,  2015, \mn@doi [\apjs]
  {10.1088/0067-0049/219/2/19}, \href
  {https://ui.adsabs.harvard.edu/abs/2015ApJS..219...19L} {219, 19}

\bibitem[\protect\citeauthoryear{{Lindegren} et~al.,}{{Lindegren}
  et~al.}{2018}]{gaiaDR2-ArXiV-2}
{Lindegren} L.,  et~al., 2018, preprint, \href
  {http://adsabs.harvard.edu/abs/2018arXiv180409366L} {} (\mn@eprint {arXiv}
  {1804.09366})

\bibitem[\protect\citeauthoryear{{Lindegren} et~al.,}{{Lindegren}
  et~al.}{2020}]{lindegrenetal20-1}
{Lindegren} L.,  et~al., 2020, arXiv e-prints, \href
  {https://ui.adsabs.harvard.edu/abs/2020arXiv201203380L} {p. arXiv:2012.03380}

\bibitem[\protect\citeauthoryear{{L{\'o}pez-Sanjuan}
  et~al.,}{{L{\'o}pez-Sanjuan} et~al.}{2019}]{lopez19}
{L{\'o}pez-Sanjuan} C.,  et~al., 2019, \mn@doi [\aap]
  {10.1051/0004-6361/201936405}, \href
  {https://ui.adsabs.harvard.edu/abs/2019A&A...631A.119L} {631, A119}

\bibitem[\protect\citeauthoryear{{Ma{\'\i}z Apell{\'a}niz} \&
  {Weiler}}{{Ma{\'\i}z Apell{\'a}niz} \& {Weiler}}{2018}]{maiz18}
{Ma{\'\i}z Apell{\'a}niz} J.,  {Weiler} M.,  2018, \mn@doi [\aap]
  {10.1051/0004-6361/201834051}, \href
  {https://ui.adsabs.harvard.edu/abs/2018A&A...619A.180M} {619, A180}

\bibitem[\protect\citeauthoryear{{McCleery} et~al.,}{{McCleery}
  et~al.}{2020}]{mccleery20}
{McCleery} J.,  et~al., 2020, \mn@doi [\mnras] {10.1093/mnras/staa2030}, \href
  {https://ui.adsabs.harvard.edu/abs/2020MNRAS.499.1890M} {499, 1890}

\bibitem[\protect\citeauthoryear{{Melis}, {Klein}, {Doyle}, {Weinberger},
  {Zuckerman}  \& {Dufour}}{{Melis} et~al.}{2020}]{melisetal20-1}
{Melis} C.,  {Klein} B.,  {Doyle} A.~E.,  {Weinberger} A.~J.,  {Zuckerman} B.,
   {Dufour} P.,  2020, arXiv e-prints, \href
  {https://ui.adsabs.harvard.edu/abs/2020arXiv201003695M} {p. arXiv:2010.03695}

\bibitem[\protect\citeauthoryear{{Mukadam} et~al.,}{{Mukadam}
  et~al.}{2013}]{mukadametal13-1}
{Mukadam} A.~S.,  et~al., 2013, \mn@doi [\apj] {10.1088/0004-637X/771/1/17},
  \href {https://ui.adsabs.harvard.edu/abs/2013ApJ...771...17M} {771, 17}

\bibitem[\protect\citeauthoryear{{Narayan} et~al.,}{{Narayan}
  et~al.}{2019}]{narayan19}
{Narayan} G.,  et~al., 2019, \mn@doi [\apjs] {10.3847/1538-4365/ab0557}, \href
  {https://ui.adsabs.harvard.edu/abs/2019ApJS..241...20N} {241, 20}

\bibitem[\protect\citeauthoryear{{Ourique}, {Romero}, {Kepler}, {Koester}  \&
  {Amaral}}{{Ourique} et~al.}{2019}]{ouriqueetal19-1}
{Ourique} G.,  {Romero} A.~D.,  {Kepler} S.~O.,  {Koester} D.,   {Amaral}
  L.~A.,  2019, \mn@doi [\mnras] {10.1093/mnras/sty2751}, \href
  {https://ui.adsabs.harvard.edu/abs/2019MNRAS.482..649O} {482, 649}

\bibitem[\protect\citeauthoryear{{Pala} et~al.,}{{Pala}
  et~al.}{2020}]{palaetal20-1}
{Pala} A.~F.,  et~al., 2020, \mn@doi [\mnras] {10.1093/mnras/staa764}, \href
  {https://ui.adsabs.harvard.edu/abs/2020MNRAS.494.3799P} {494, 3799}

\bibitem[\protect\citeauthoryear{{Queiroz} et~al.,}{{Queiroz}
  et~al.}{2020}]{Queiroz20}
{Queiroz} A.~B.~A.,  et~al., 2020, \mn@doi [\aap]
  {10.1051/0004-6361/201937364}, \href
  {https://ui.adsabs.harvard.edu/abs/2020A&A...638A..76Q} {638, A76}

\bibitem[\protect\citeauthoryear{{Remy}, {Grenier}, {Marshall}  \&
  {Casandjian}}{{Remy} et~al.}{2018}]{Remy18}
{Remy} Q.,  {Grenier} I.~A.,  {Marshall} D.~J.,   {Casandjian} J.~M.,  2018,
  \mn@doi [\aap] {10.1051/0004-6361/201731488}, \href
  {https://ui.adsabs.harvard.edu/abs/2018A&A...616A..71R} {616, A71}

\bibitem[\protect\citeauthoryear{{Riello} et~al.,}{{Riello}
  et~al.}{2020}]{rielloetal20-1}
{Riello} M.,  et~al., 2020, arXiv e-prints, \href
  {https://ui.adsabs.harvard.edu/abs/2020arXiv201201916R} {p. arXiv:2012.01916}

\bibitem[\protect\citeauthoryear{{Rolland}, {Bergeron}  \&
  {Fontaine}}{{Rolland} et~al.}{2018}]{rolland2018}
{Rolland} B.,  {Bergeron} P.,   {Fontaine} G.,  2018, \mn@doi [\apj]
  {10.3847/1538-4357/aab713}, \href
  {http://adsabs.harvard.edu/abs/2018ApJ...857...56R} {857, 56}

\bibitem[\protect\citeauthoryear{{Rybizki}, {Green}, {Rix}, {Demleitner},
  {Zari}, {Udalski}, {Smart}  \& {Gould}}{{Rybizki}
  et~al.}{2021}]{rybizkietal21-1}
{Rybizki} J.,  {Green} G.,  {Rix} H.-W.,  {Demleitner} M.,  {Zari} E.,
  {Udalski} A.,  {Smart} R.~L.,   {Gould} A.,  2021, arXiv e-prints, \href
  {https://ui.adsabs.harvard.edu/abs/2021arXiv210111641R} {p. arXiv:2101.11641}

\bibitem[\protect\citeauthoryear{{Sanders} \& {Das}}{{Sanders} \&
  {Das}}{2018}]{Sanders18}
{Sanders} J.~L.,  {Das} P.,  2018, \mn@doi [\mnras] {10.1093/mnras/sty2490},
  \href {https://ui.adsabs.harvard.edu/abs/2018MNRAS.481.4093S} {481, 4093}

\bibitem[\protect\citeauthoryear{{Schlafly} \& {Finkbeiner}}{{Schlafly} \&
  {Finkbeiner}}{2011}]{schlafy}
{Schlafly} E.~F.,  {Finkbeiner} D.~P.,  2011, \mn@doi [\apj]
  {10.1088/0004-637X/737/2/103}, \href
  {http://adsabs.harvard.edu/abs/2011ApJ...737..103S} {737, 103}

\bibitem[\protect\citeauthoryear{{Schlegel}, {Finkbeiner}  \&
  {Davis}}{{Schlegel} et~al.}{1998}]{schlegeletal98-1}
{Schlegel} D.~J.,  {Finkbeiner} D.~P.,   {Davis} M.,  1998, ApJ, \href
  {1998ApJ...500..525S} {500, 525}

\bibitem[\protect\citeauthoryear{{Serenelli}, {Althaus}, {Rohrmann}  \&
  {Benvenuto}}{{Serenelli} et~al.}{2001}]{serenelli01}
{Serenelli} A.~M.,  {Althaus} L.~G.,  {Rohrmann} R.~D.,   {Benvenuto} O.~G.,
  2001, \mn@doi [\mnras] {10.1046/j.1365-8711.2001.04449.x}, \href
  {https://ui.adsabs.harvard.edu/abs/2001MNRAS.325..607S} {325, 607}

\bibitem[\protect\citeauthoryear{{Sion}, {Greenstein}, {Landstreet}, {Liebert},
  {Shipman}  \& {Wegner}}{{Sion} et~al.}{1983}]{sionetal83-1}
{Sion} E.~M.,  {Greenstein} J.~L.,  {Landstreet} J.~D.,  {Liebert} J.,
  {Shipman} H.~L.,   {Wegner} G.~A.,  1983, \mn@doi [ApJ] {10.1086/161036},
  \href {http://adsabs.harvard.edu/abs/1983ApJ...269..253S} {269, 253}

\bibitem[\protect\citeauthoryear{{Sion}, {Holberg}, {Oswalt}, {McCook},
  {Wasatonic}  \& {Myszka}}{{Sion} et~al.}{2014}]{sionetal14-1}
{Sion} E.~M.,  {Holberg} J.~B.,  {Oswalt} T.~D.,  {McCook} G.~P.,  {Wasatonic}
  R.,   {Myszka} J.,  2014, \mn@doi [AJ] {10.1088/0004-6256/147/6/129}, \href
  {http://adsabs.harvard.edu/abs/2014AJ....147..129S} {147, 129}

\bibitem[\protect\citeauthoryear{{Tian} et~al.,}{{Tian}
  et~al.}{2017}]{tianetal17-1}
{Tian} H.-J.,  et~al., 2017, \mn@doi [ApJS] {10.3847/1538-4365/aa826a}, \href
  {http://adsabs.harvard.edu/abs/2017ApJS..232....4T} {232, 4}

\bibitem[\protect\citeauthoryear{{Torres}, {Cantero}, {Rebassa-Mansergas},
  {Skorobogatov}, {Jim{\'e}nez-Esteban}  \& {Solano}}{{Torres}
  et~al.}{2019}]{torresetal19-1}
{Torres} S.,  {Cantero} C.,  {Rebassa-Mansergas} A.,  {Skorobogatov} G.,
  {Jim{\'e}nez-Esteban} F.~M.,   {Solano} E.,  2019, \mn@doi [\mnras]
  {10.1093/mnras/stz814}, \href
  {https://ui.adsabs.harvard.edu/abs/2019MNRAS.485.5573T} {485, 5573}

\bibitem[\protect\citeauthoryear{{Torres}, {Rebassa-Mansergas}, {Camisassa}  \&
  {Raddi}}{{Torres} et~al.}{2021}]{torresetal21-1}
{Torres} S.,  {Rebassa-Mansergas} A.,  {Camisassa} M.~E.,   {Raddi} R.,  2021,
  \mn@doi [\mnras] {10.1093/mnras/stab079}, \href
  {https://ui.adsabs.harvard.edu/abs/2021MNRAS.502.1753T} {502, 1753}

\bibitem[\protect\citeauthoryear{{Tremblay}, {Bergeron}  \&
  {Gianninas}}{{Tremblay} et~al.}{2011}]{tremblayetal11-1}
{Tremblay} P.-E.,  {Bergeron} P.,   {Gianninas} A.,  2011, \mn@doi [ApJ]
  {10.1088/0004-637X/730/2/128}, \href
  {http://adsabs.harvard.edu/abs/2011ApJ...730..128T} {730, 128}

\bibitem[\protect\citeauthoryear{{Tremblay}, {Kalirai}, {Soderblom}, {Cignoni}
  \& {Cummings}}{{Tremblay} et~al.}{2014}]{tremblayetal14-1}
{Tremblay} P.-E.,  {Kalirai} J.~S.,  {Soderblom} D.~R.,  {Cignoni} M.,
  {Cummings} J.,  2014, \mn@doi [\apj] {10.1088/0004-637X/791/2/92}, \href
  {http://adsabs.harvard.edu/abs/2014ApJ...791...92T} {791, 92}

\bibitem[\protect\citeauthoryear{{Tremblay}, {Cukanovaite}, {Gentile Fusillo},
  {Cunningham}  \& {Hollands}}{{Tremblay} et~al.}{2019a}]{tremblay2019}
{Tremblay} P.-E.,  {Cukanovaite} E.,  {Gentile Fusillo} N.~P.,  {Cunningham}
  T.,   {Hollands} M.~A.,  2019a, \mn@doi [\mnras] {10.1093/mnras/sty3067},
  \href {http://adsabs.harvard.edu/abs/2019MNRAS.482.5222T} {482, 5222}

\bibitem[\protect\citeauthoryear{{Tremblay} et~al.,}{{Tremblay}
  et~al.}{2019b}]{tremblay-nature}
{Tremblay} P.-E.,  et~al., 2019b, \mn@doi [\nat] {10.1038/s41586-018-0791-x},
  \href {https://ui.adsabs.harvard.edu/abs/2019Natur.565..202T} {565, 202}

\bibitem[\protect\citeauthoryear{{Tremblay} et~al.,}{{Tremblay}
  et~al.}{2020}]{tremblay20}
{Tremblay} P.~E.,  et~al., 2020, \mn@doi [\mnras] {10.1093/mnras/staa1892},
  \href {https://ui.adsabs.harvard.edu/abs/2020MNRAS.497..130T} {497, 130}

\bibitem[\protect\citeauthoryear{{Vergely}, {Valette}, {Lallement}  \&
  {Raimond}}{{Vergely} et~al.}{2010}]{Vergely10}
{Vergely} J.-L.,  {Valette} B.,  {Lallement} R.,   {Raimond} S.,  2010, \mn@doi
  [A\&A] {10.1051/0004-6361/200913962}, \href
  {http://adsabs.harvard.edu/abs/2010A%26A...518A..31V} {518, A31}

\bibitem[\protect\citeauthoryear{{Zuckerman}, {Becklin}, {Macintosh}  \&
  {Bida}}{{Zuckerman} et~al.}{1997}]{zuckermanetal97-01}
{Zuckerman} B.,  {Becklin} E.~E.,  {Macintosh} B.~A.,   {Bida} T.,  1997,
  \mn@doi [\aj] {10.1086/118296}, \href
  {https://ui.adsabs.harvard.edu/abs/1997AJ....113..764Z} {113, 764}

\bibitem[\protect\citeauthoryear{{de Jong} et~al.,}{{de Jong}
  et~al.}{2014}]{4most14-1}
{de Jong} R.~S.,  et~al., 2014, in Society of Photo-Optical Instrumentation
  Engineers (SPIE) Conference Series. p.~0, \mn@doi{10.1117/12.2055826}

\bibitem[\protect\citeauthoryear{{van Horn}}{{van Horn}}{1968}]{vanhorne68-1}
{van Horn} H.~M.,  1968, \mn@doi [\apj] {10.1086/149432}, \href
  {https://ui.adsabs.harvard.edu/abs/1968ApJ...151..227V} {151, 227}

\makeatother
\end{thebibliography}


\appendix

\section{Changes in the the northern hemisphere 40pc sample}

\begin{table*}
\scriptsize
\caption{Changes in the northern 40\,pc white dwarf sample between DR2 and EDR3. This excludes 519 spectroscopically confirmed white dwarfs that are part of the sample in both DR2 and EDR3 (see table A1 of \citealt{mccleery20}).}
\label{tab:appendix_40pc}
\begin{tabular}{llllllll}
\hline
Gaia DR3 ID & WD name & P$_{\rm WD}$ & RA & DEC & $\varpi (\sigma_{\varpi})$ & $G$ $(\sigma_{\rm G})$ & Comment \\
 & (confirmed WDs only) & & (deg) & (deg) & (mas) & (mag) &\\
\hline
\multicolumn{8}{l}{A) New EDR3 catalogue entries}\\
\hline
2545505281002947200 & WD 0011+000             & 0.995 & 3.415   & 0.322  & 25.002 (0.041) & 15.336 (0.002) & SpT: DA \citep{limoges15}, moved to 40\,pc\\
307323228064848512  & WD 0108+277             & 0.748 & 17.685  & 27.970 & 26.312 (0.082) & 16.153 (0.003) & SpT: DAZ \citep{kawka06}, incomplete entry in DR2\\
3320184202856435840 & WD 0553+053             & 0.995 & 89.104  & 5.359  & 123.198 (0.017)& 13.969 (0.002) & SpT: DAH \citep{limoges15}, incomplete entry in DR2\\
1146403741412820864 & WDJ102203.66+824310.00  & 0.991 & 155.505 & 82.718 & 25.113 (0.092) & 17.899 (0.003) & SpT: DA \citep{tremblay20}, moved to 40\,pc\\
3883918657822146944 & --                      & 0.994 & 157.361 & 12.959 & 27.751 (0.166) & 17.460 (0.003) & New EDR3 white dwarf candidate\\
3978879594463300992 & WD 1121+216             & 0.992 & 171.049 & 21.359 & 68.041 (0.028) & 14.124 (0.002) & SpT: DA \citep{limoges15}, incomplete entry in DR2\\
3920187251456610816 & WD 1153+135             & 0.977 & 179.048 & 13.265 & 28.115 (0.090) & 17.401 (0.002) & SpT: DC \citep{leggett18}, incomplete entry in DR2\\
3905335598144227200 & WD 1218+095             & 1.000 & 185.201 & 9.235  & 26.681 (0.346) & 19.606 (0.005) & SpT: DC \citep{leggett18} incomplete entry in DR2\\
3713594960831605760 & WD 1334+039             & 0.996 & 204.116 & 3.674  & 119.756 (0.030)& 14.376 (0.002) & SpT: DA \citep{limoges15}, missing from DR2\\
1336988963803208192 & --                      & 0.941 & 259.981 & 36.657 & 28.539 (0.051) & 16.953 (0.003) & Suspicious candidate, 4'' wide companion to M dwarf\\
4512265810525783680 & --                      & 0.993 & 281.752 & 18.185 & 34.486 (0.037) & 15.332 (0.002) & Suspicious candidate, 20'' wide companion to HD 173880\\
4539227892919675648 & WDJ184733.18+282057.54  & 0.999 & 281.889 & 28.347 & 25.001 (0.120) & 18.439 (0.003) & SpT: DC \citep{tremblay20}, moved to 40\,pc\\
4288942973032203904 & --                      & 1.000 & 290.526 & 2.553  & 25.377 (0.265) & 19.119 (0.004) & SpT: DZ \citep{tremblay20}, moved to 40\,pc\\
2701893698904233216 & LP 578-24               & 0.986 & 325.726 & 8.090  & 33.449 (0.086) & 17.052 (0.002) & SpT: DA \citep{limoges15}, incomplete entry in DR2\\
2730707260103011712 & WD 2220+121.1           & 0.999 & 335.644 & 12.362 & 25.996 (0.218) & 18.366 (0.003) & SpT: DC \citep{kilicetal06-3}, incomplete entry in DR2\\
\hline
\multicolumn{8}{l}{B) DR2 white dwarf members moved out of the sample}\\
\hline
3127761765259717632 & WDJ065722.88+024100.84  & 0.997 & 104.345 & 2.682  & 24.879 (0.056) & 16.096 (0.02) & SpT: DC \citep{limoges15}\\
3860381618565361024 & WDJ102459.83+044610.50  & 0.997 & 156.248 & 4.769  & 23.204 (0.046) & 14.214 (0.02) & SpT: DA \citep{gianninas2011}\\  
\hline
\multicolumn{8}{l}{C) Confirmed northern 40\,pc white dwarfs missing from DR2 and EDR3}\\
\hline
975968340912517248  & WD 0727+482A            & --    & 112.695 & 48.168 & --             & 15.063 (0.02) & SpT: DA \citep{limoges15},\\
 & & & & & & &
$\varpi=88.723\pm 0.029$ (Gaia EDR3 companion G 107-69)\\
975968340910692608  & WD 0727+482B            & --    & 112.695 & 48.168 & --             & 15.251 (0.02) & SpT: DA \citep{limoges15},\\
 & & & & & & & $\varpi=88.723\pm 0.029$ (Gaia EDR3 companion G 107-69)\\
--                  & WD 0736+053             & --    & --      & --     & --             & --            & SpT: DQZ \citep{limoges15},\\
 & & & & & & & $\varpi=284.56 \pm 1.26$ (van Leeuwen 2007)\\
3817534337626005632 & WD 1120+073             & --    & 170.881 & 7.022  & --             & 17.511 (0.02) & SpT: DC \citep{limoges15},\\
 & & & & & & & $\varpi=27.025\pm 0.469$ (Gaia EDR3 companion LP 552-48)\\
3701290326205270528 & WD 1214+032             & --    & 184.213 & 2.968  & 42.772 (0.042) & 15.330 (0.02) & SpT: DA \citep{limoges15},\\
 & & & & & & & no colour information in EDR3\\
1362295082910739840 & HD 159062B              & --    & 262.568 & 47.402 & --             & 16.745 (0.02) & SpT: G9V+WD \citep{hirsch19},\\
 & & & & & & & $\varpi=46.185\pm 0.014$ (Gaia EDR3 companion HD 159062)\\
2274076301516712704 & WD 2126+734B            & --    & 321.741 & 73.643 & 44.909 (0.069) & 16.511 (0.02) & SpT: DC \citep{zuckermanetal97-01},\\
 & & & & & & & missing from catalogue due to large BP/RP excess factor\\
1962707287281651712 & PM J22105+4532          & --    & 332.643 & 45.542 & 27.858 (0.078) & 17.146 (0.02) & SpT: DC \citep{limoges15}, no colour information in EDR3\\
\hline
\multicolumn{8}{l}{D) Unresolved binaries missing from our catalogues}\\
\hline
1005873614080407296 & LHS 1817                & --    & 91.375  & 60.819 & 61.426 (0.053) & 12.296 (0.03) & SpT: M4.5V+WD \citep{mccleery20}\\
3845263368043086080 & WD 0911+023             & --    & 138.591 & 2.312  & 27.070 (0.616) & 3.88 (0.05)   & SpT: B9.5V+WD \citep{mccleery20}\\
1548104507825815296 & WD 1213+528             & --    & 183.933 & 52.516 & 34.949 (0.021) & 12.570 (0.03) & SpT: DA+dM \citep{mccleery20}\\
4478524169500496000 & HD 169889               & --    & 276.590 & 8.613  & 28.279 (0.025) & 8.10 (0.03)   & SpT: G9V+WD \citep{mccleery20}\\
1550299304833675392 & WD 1324+458             & --    & 201.720 & 45.545 & 32.772 (0.021) & 11.962 (0.03) & SpT: M3V+DA \citep{mccleery20}\\
\hline
\multicolumn{8}{l}{E) EDR3 and DR2 catalogue entries without spectral confirmation}\\
\hline
283928743068277376  & WDJ050600.41+590326.89  &  1.000 & 76.501  & 59.055 & 27.731 (0.332) & 19.635 (0.004) & Unobserved high $P_{\rm WD}$ DR2 candidate (table A1 of \citealt{mccleery20})\\
3346787883122375680 & WDJ055602.01+135446.71  &	 0.996 & 89.0100 & 13.910 & 36.531 (0.084) & 16.928 (0.002) & Unobserved high $P_{\rm WD}$ DR2 candidate (table A1 of \citealt{mccleery20})\\
611074413433751680  & WDJ090834.39+172148.53  &  0.914 & 137.142 & 17.362 & 30.676 (0.058) & 16.591 (0.002) & Unobserved low $P_{\rm WD}$ DR2 candidate (table A2 of \citealt{mccleery20})\\
3982007636324256000 & WDJ110143.04+172139.39  &	 0.994 & 165.427 & 17.359 & 34.668 (0.053) & 15.973 (0.003) & Unobserved high $P_{\rm WD}$ DR2 candidate (table A1 of \citealt{mccleery20})\\
4004185576130620288 & WDJ115007.08+240403.54  &  1.000 & 177.526 & 24.064 & 33.198 (0.284) & 19.357 (0.004) & Unobserved low $P_{\rm WD}$ DR2 candidate (table A2 of \citealt{mccleery20})\\
1688618481786030336 & WDJ131830.01+735318.25  &  0.964 & 199.622 & 73.888 & 27.448 (0.138) & 16.864 (0.005) & Unobserved low $P_{\rm WD}$ DR2 candidate (table A2 of \citealt{mccleery20})\\
2149331587745863680 & WDJ181548.96+553232.22  &  0.747 & 273.954 & 55.541 & 26.365 (0.051) & 17.098 (0.002) & Unobserved low $P_{\rm WD}$ DR2 candidate (table A2 of \citealt{mccleery20})\\
2127093140445053696 & WDJ191936.23+452743.55  &  0.866 & 289.900 & 45.463 & 35.639 (0.035) & 16.438 (0.002) & Unobserved low $P_{\rm WD}$ DR2 candidate (table A2 of \citealt{mccleery20})\\
\hline
\end{tabular}
\end{table*}


\bsp	
\label{lastpage}
\end{document}